\documentclass[acmtog]{acmart}

\usepackage{booktabs} 

\usepackage{subcaption}
\usepackage{ifplatform}
\usepackage{epstopdf}
\usepackage{flushend}

\usepackage[ruled]{algorithm2e} 

\SetAlFnt{\small}
\SetAlCapFnt{\small}
\SetAlCapNameFnt{\small}
\SetAlCapHSkip{0pt}

\setcopyright{rightsretained}

\newcommand{\mydraft}{false}

\newcommand{\mb}{\mathbf}
\newcommand{\xx}{\mb{x}}

\newcommand{\bb}{\mb{b}}

\newcommand{\cc}{\mb{c}}

\newcommand{\uu}{\mb{u}}
\ifwindows
\renewcommand\vv{\mb{v}}
\else
\newcommand{\vv}{\mb{v}}
\fi

\newcommand{\MM}{\mb{M}}

\newcommand{\PP}{\mb{P}}

\newcommand{\rr}{\mb{r}}

\newcommand{\II}{\mb{I}}

\newcommand{\BB}{\mb{B}}
\newcommand{\UU}{\mb{U}}

\newcommand{\GG}{\mb{G}}

\newcommand{\ii}{\mb{i}}
\newcommand{\jj}{\mb{j}}

\newcommand{\WW}{\mb{W}}

\newcommand{\ww}{\mb{w}}

\newcommand{\dd}{\mb{d}}
\newcommand{\DD}{\mb{D}}
\newcommand{\nn}{\mb{n}}
\renewcommand{\ll}{\mb{l}}

\renewcommand{\gg}{\mb{g}}

\renewcommand{\AA}{\mb{A}}

\newcommand{\RR}{\mb{R}}

\newcommand{\subf}[1]{%
  {\small\begin{tabular}[t]{@{}c@{}}
  #1
  \end{tabular}}%
}

\begin{document}
\title{A Hybrid Lagrangian/Eulerian Collocated Advection and Projection Method for Fluid Simulation}


\author{Steven W. Gagniere}
\affiliation{%
  \institution{UCLA}}
\email{sgagniere@math.ucla.edu}

\author{David A. B. Hyde}
\affiliation{%
  \institution{UCLA}}
\email{dabh@math.ucla.edu}

\author{Alan Marquez-Razon}
\affiliation{%
  \institution{UCLA}}
\email{marqueza04@g.ucla.edu}

\author{Chenfanfu Jiang}
\affiliation{%
  \institution{University of Pennsylvania}}
\email{cffjiang@seas.upenn.edu}

\author{Ziheng Ge}
\affiliation{%
  \institution{UCLA}}
\email{zihengge@math.ucla.edu}

\author{Xuchen Han}
\affiliation{%
  \institution{UCLA}}
\email{xhan0619@ucla.edu}

\author{Qi Guo}
\affiliation{%
  \institution{UCLA}}
\email{qiguo@ucla.edu}

\author{Joseph Teran}
\affiliation{%
  \institution{UCLA}}
\email{jteran@math.ucla.edu}

\begin{teaserfigure}
\centering
\includegraphics[draft=\mydraft,trim={0 0 0 0},clip,width=1.0\textwidth]{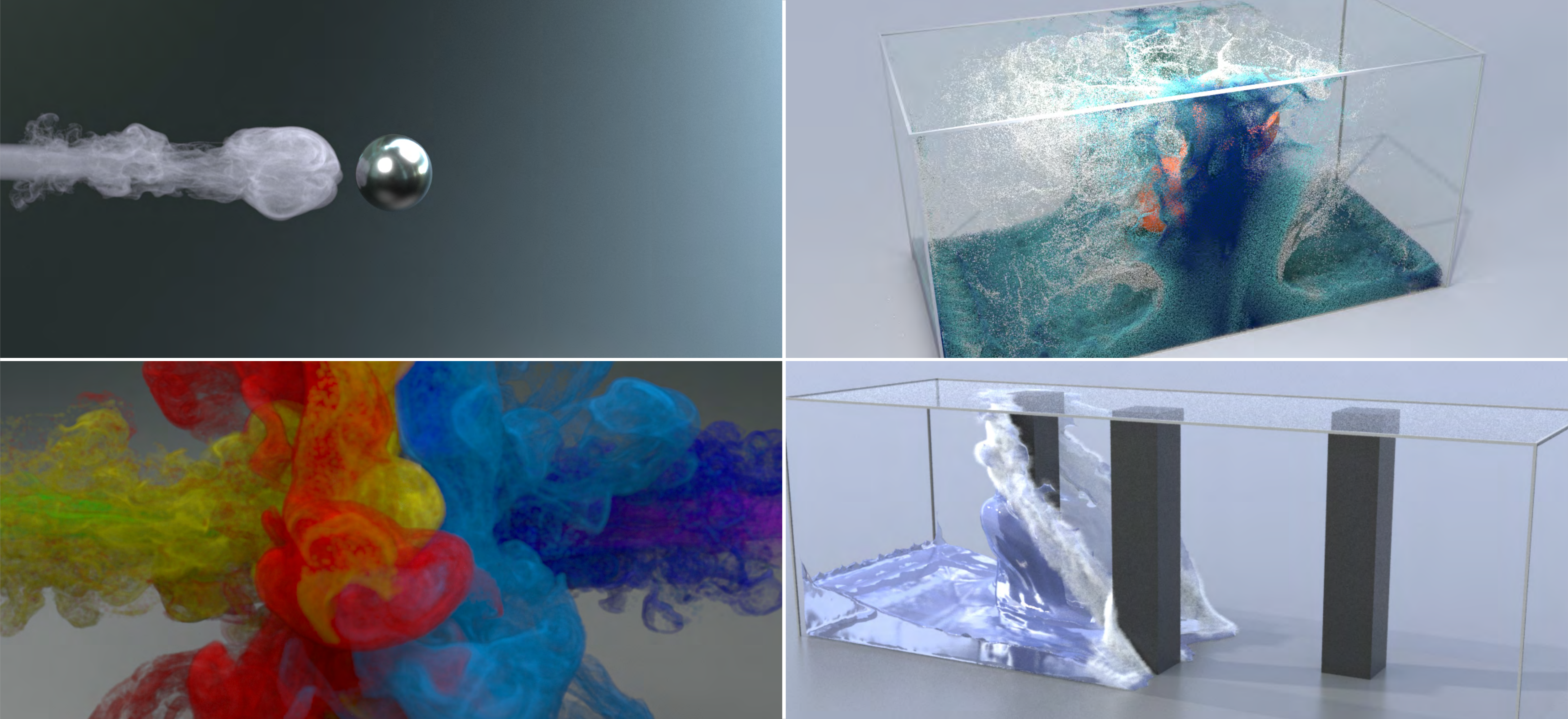}
\caption{We simulate detailed incompressible flows with free surfaces and irregular domains using our novel hybrid particle/grid simulation approach.  Our numerical method yields intricate flow details with little dissipation, even at modest spatial resolution. Furthermore, we use collocated rather than staggered MAC grids.}\label{fig:hero}
\end{teaserfigure}

\begin{abstract}
We present a hybrid particle/grid approach for simulating incompressible fluids on collocated velocity grids. We interchangeably use particle and grid representations of transported quantities to balance efficiency and accuracy. A novel Backward Semi-Lagrangian method is derived to improve accuracy of grid based advection. Our approach utilizes the implicit formula associated with solutions of Burgers' equation. We solve this equation using Newton's method enabled by $C^1$ continuous grid interpolation. We enforce incompressibility over collocated, rather than staggered grids. Our projection technique is variational and designed for B-spline interpolation over regular grids where multiquadratic interpolation is used for velocity and multilinear interpolation for pressure. Despite our use of regular grids, we extend the variational technique to allow for cut-cell definition of irregular flow domains for both Dirichlet and free surface boundary conditions.
\end{abstract}

%
%


%
%


\maketitle

\section{Introduction}

\begin{figure}[!ht]
  \centering
  \begin{subfigure}{1.0\columnwidth}
     \includegraphics[draft=\mydraft,trim={0 200px 0 200px},clip,width=\columnwidth]{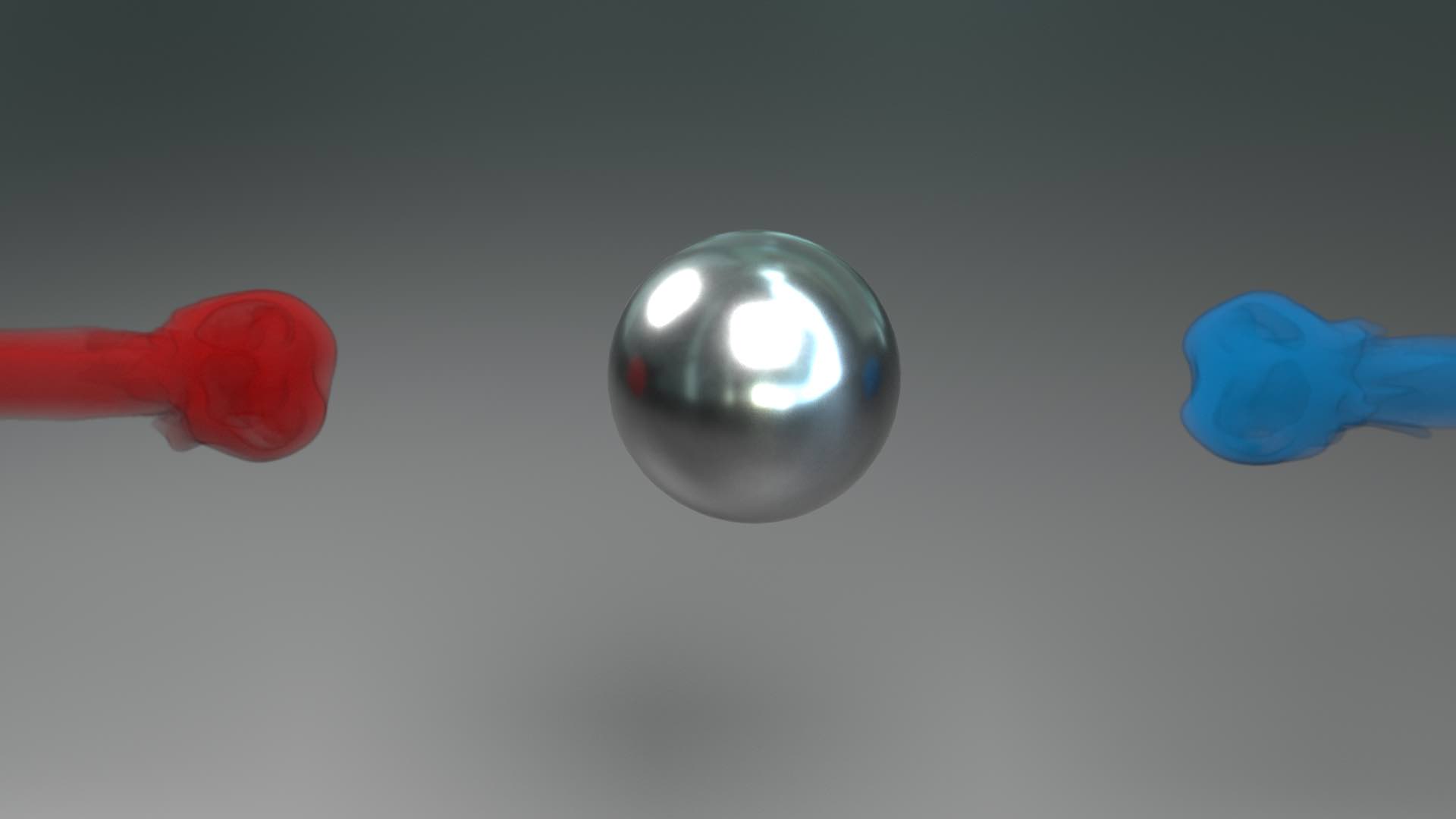}
  \end{subfigure} \\
  \begin{subfigure}{1.0\columnwidth}
     \includegraphics[draft=\mydraft,trim={0 200px 0 200px},clip,width=\columnwidth]{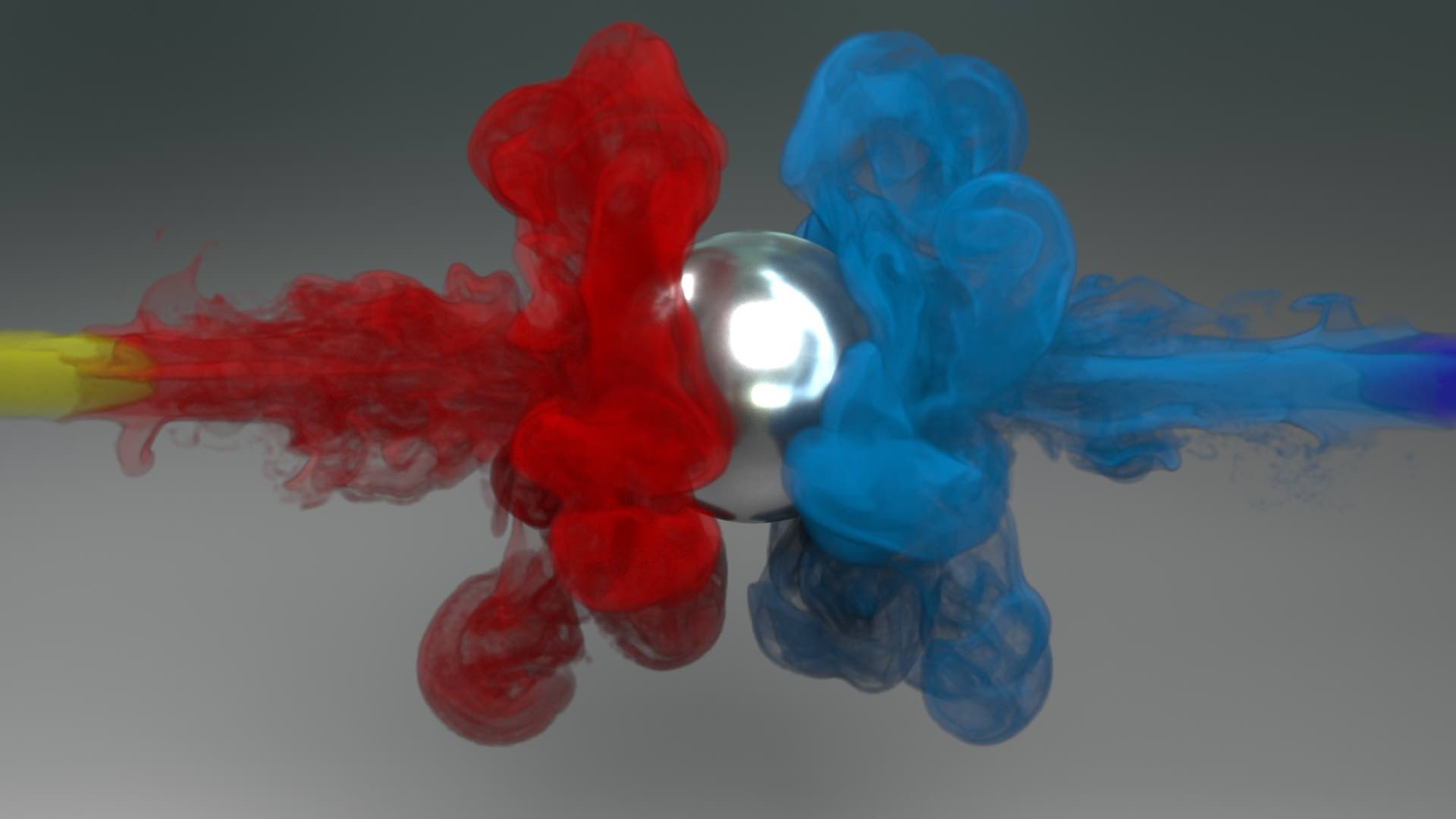}
  \end{subfigure} \\
  \begin{subfigure}{1.0\columnwidth}
     \includegraphics[draft=\mydraft,trim={0 200px 0 200px},clip,width=\columnwidth]{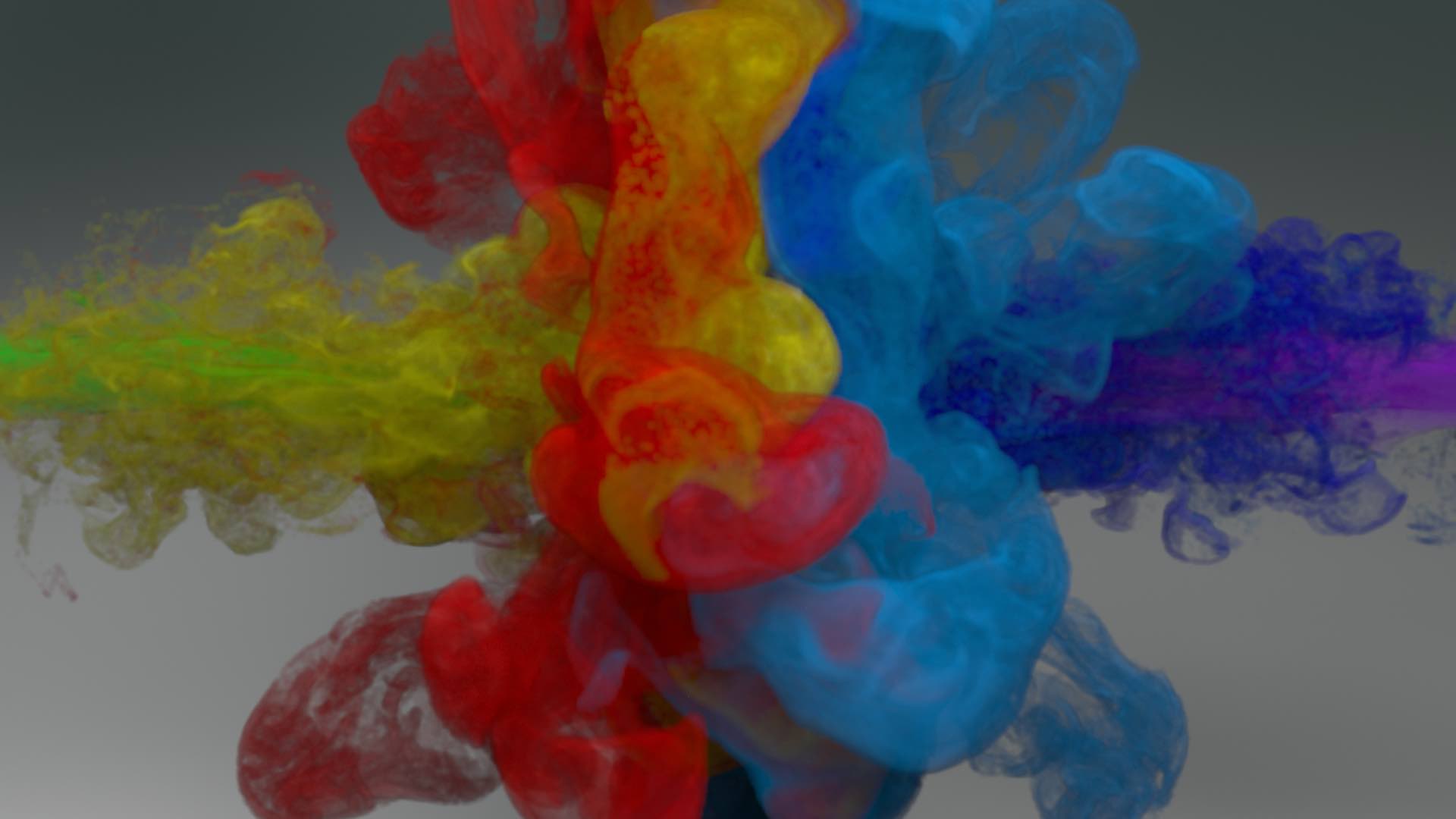}
  \end{subfigure} 
\caption{{\textbf{Colorful smoke jets}}. Multicolored jets of smoke are simulated with BSLQB. Intricate mixing is induced as the flows collide at the spherical boundary.}\label{fig:rainbowsmoke}
\end{figure}

Whether it be billowing smoke, energetic explosions, or breaking waves, simulation of incompressible flow is an indispensable tool for modern visual effects. Ever since the pioneering works of Foster and Metaxas \shortcite{foster:1996:liquids}, Stam \shortcite{stam:1999:stable} and Fedkiw et al. \shortcite{fedkiw:2001:visual,foster:2001:practical}, the Chorin \shortcite{chorin:1967:numerical} splitting of advective and pressure projection terms  has been the standard in computer graphics applications \cite{bridson:2008:fluid-simulation}. Most techniques use regular grids of Marker-And-Cell (MAC) \cite{harlow:1965:viscous-flow} type with pressure and velocity components staggered at cell centers and faces respectively. Furthermore, advection is most often discretized using semi-Lagrangian techniques originally developed in the atmospheric sciences \cite{stam:1999:stable,robert:1981:stable}. Although well-established, these techniques are not without their drawbacks. For example, the staggering utilized in the MAC grids is cumbersome since variables effectively live on four different grids. This can complicate many algorithms related to incompressible flow. E.g. Particle-In-Cell (PIC) \cite{harlow:1964:pic} techniques like FLIP \cite{brackbill:1986:flip-pic,zhu:2005:sand-fluid}, Affine/Polynomial Particle-In-Cell (APIC/PolyPIC) \cite{jiang:2015:apic,fu:2017:poly} and the Material Point Method (MPM) \cite{sulsky:1994:history-materials,stomakhin:2014:augmented-mpm} must transfer information separately to and from each individual grid. Similarly, semi-Lagrangian techniques must separately solve for upwind locations at points on each of the velocity component grids. Moreover, while semi-Lagrangian techniques are renowned for the large time steps they admit (notably larger than the Courant-Friedrichs-Levy (CFL) condition), their inherent stability is plagued by dissipation that must be removed for most visual effects phenomena. Another limitation of the MAC grid arises with free-surface water simulation. In this case, the staggering prevents many velocity components near the fluid free surface from receiving a correction during projection (see e.g. \cite{bridson:2008:fluid-simulation}). Each of these velocity components must then be separately extrapolated to from the interior to receive a pressure correction. \\
\\
MAC grids are useful because the staggering prevents pressure null modes while allowing for accurate second order central differencing in discrete grad/div operators. However, there are alternatives in the computational physics literature. Many mixed Finite Element Method (FEM) techniques use collocated velocities \cite{hughes:2000:book} without suffering from pressure mode instabilities. For example, Taylor-Hood elements \cite{taylor:1973:TH} use collocated multi-quadratic velocity interpolation and multilinear pressure interpolation to enforce incompressiblity. Recently, B-spline interpolation \cite{deboor:1978:splines} has been used with Taylor-Hood \cite{bressan:2010:isogeometric}. We build on this work and develop an approach based on collocated multi-quadratic B-spline interpolation for velocities. This choice is motivated by the simplicity of collocated grids compared to staggering, but also because of the ease of attaining continuous derivatives with B-spline interpolation. For example, this interpolation is often chosen with MPM applications since $C^1$ interpolation is essential for stability \cite{steffen:2008:analysis}. In the context of fluids, we show that this allows for extremely stable and accurate advection.
\\
\\
We develop a new approach for Chorin splitting \shortcite{chorin:1967:numerical} based on the collocated multiquadratic B-spline velocity, multilinear pressure Taylor-Hood element \cite{bressan:2010:isogeometric}. However, unlike the fully collocated technique of Bressan \shortcite{bressan:2010:isogeometric}, we stagger pressures on the nodes of the grid and velocities at cell centers as in \cite{ando:2013:surfacing}, since it reduces coupling in the pressure projection system and naturally accommodates particle-based definition of the flow domain for free-surface simulation of water. Notably, our formulation does not require velocity extrapolation after pressure projection for free-surface flow calculations as is typically needed with MAC grids. We use regular grids, but as in \cite{batty:2007:solid-fluid,batty:2008:buckling,larionov:2017:stokes}, we allow for irregular domains in a variational way using cut cells. However, rather than a weighted finite different approach, we use an FEM approach as in XFEM \cite{belytschko:2009:review,koschier:2017:xfem} and virtual node (VNA) \cite{schroeder:2014:vna} techniques. In VNA and XFEM approaches, integrals arising in the variational formulation are carried out over the intersection of the grid with the domain geometry. \\
\\
\begin{figure}[!ht]
  \centering
  \begin{subfigure}{\columnwidth}
     \includegraphics[draft=\mydraft,trim={0 100px 0 50px},clip,width=\columnwidth]{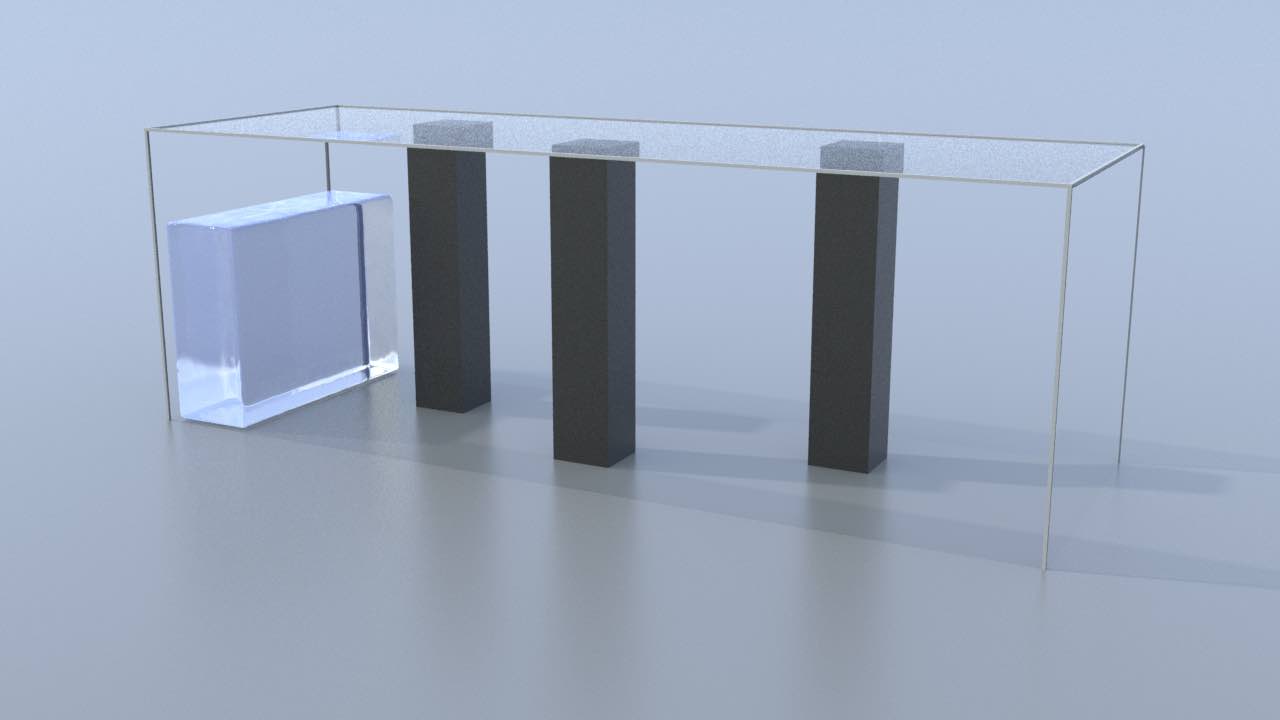}
  \end{subfigure} \\ 
  \begin{subfigure}{\columnwidth}
     \includegraphics[draft=\mydraft,trim={0 100px 0 50px},clip,width=\columnwidth]{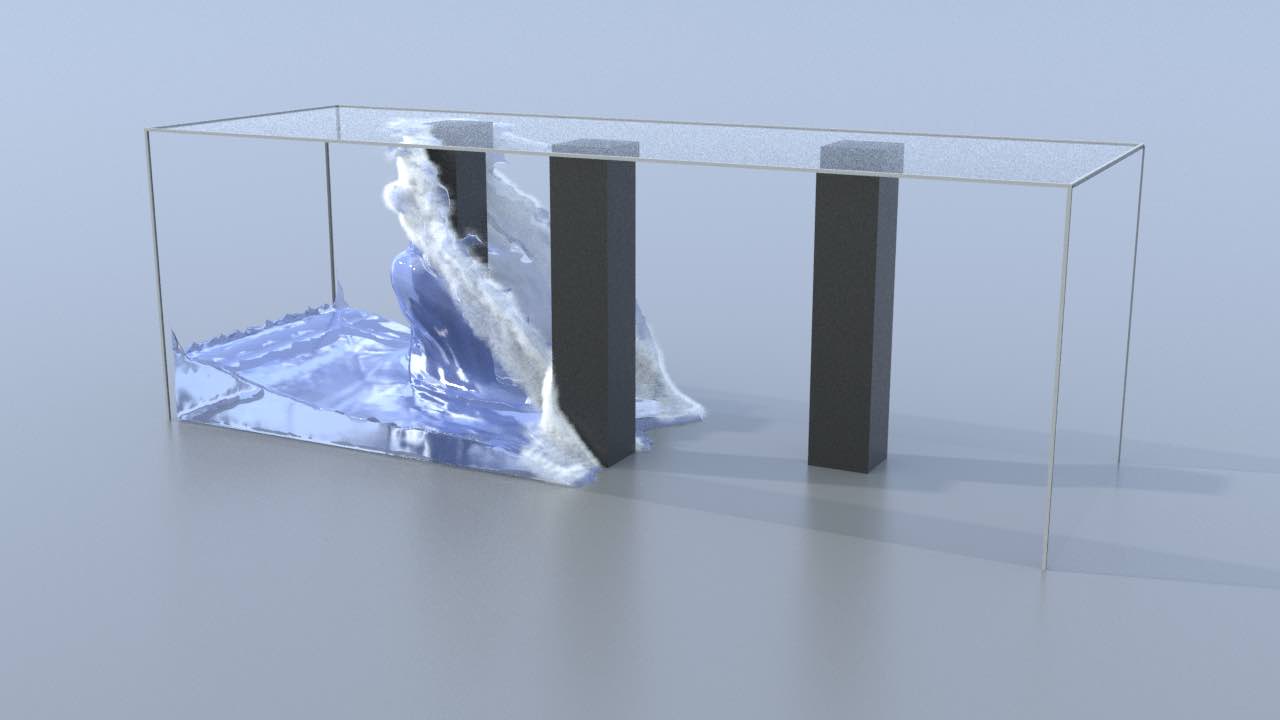}
  \end{subfigure} \\ 
  \begin{subfigure}{\columnwidth}
     \includegraphics[draft=\mydraft,trim={0 100px 0 50px},clip,width=\columnwidth]{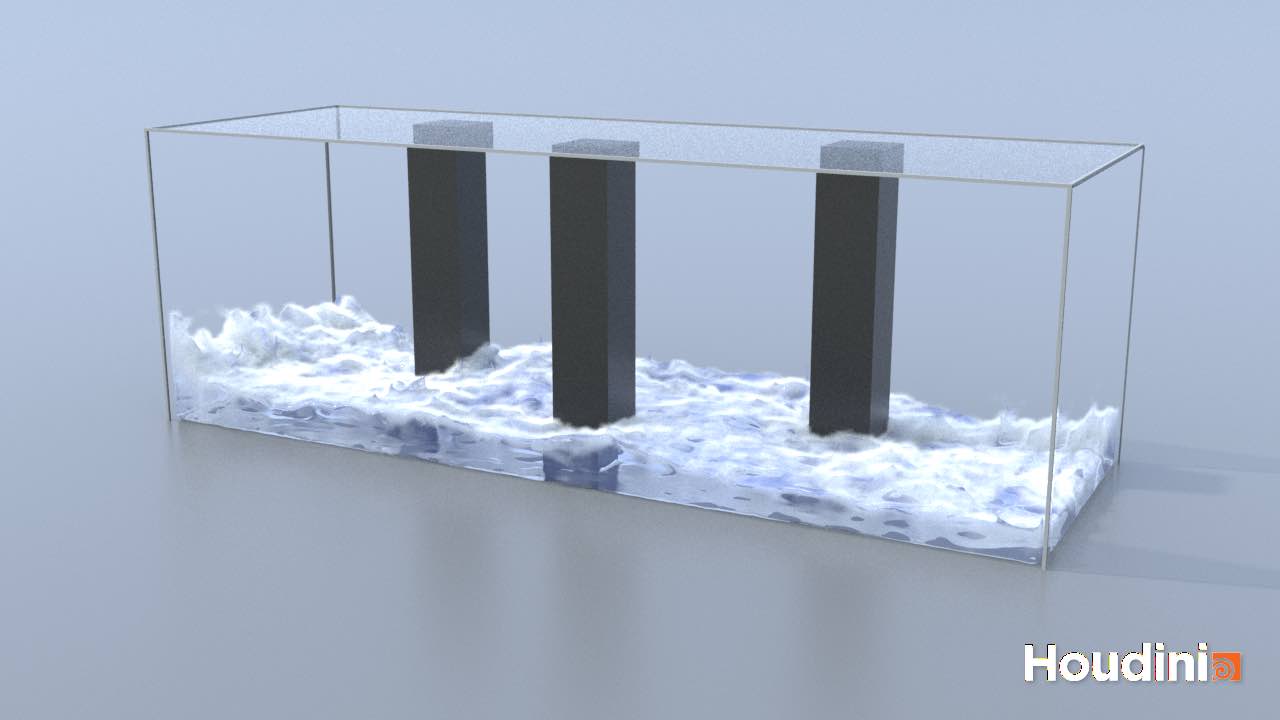}
  \end{subfigure}
\caption{{\textbf{Dam break}}. A block of water falls in a rectangular domain with obstacles.  Dynamic splashing behavior is followed by settling of the water in the tank. White water rendering effects are added based on \cite{ihmsen:2012:unified}.}\label{fig:dambreak_final}
\end{figure}
We leverage $C^1$ continuity guaranteed by our quadratic B-spline velocity interpolation to develop BSLQB, a novel Backward Semi-Lagrangian (BSL) \cite{robert:1981:stable} technique that achieves second order accuracy in space and time. BSL techniques utilize the implicit form of semi-Lagrangian advection. We show that our novel BSL method for quadratic B-splines dramatically reduces numerical dissipation with only a small modification to the widely-adopted explicit semi-Lagrangian formulations typically used in graphics applications. Semi-Lagrangian techniques for velocity advection utilize the implicit relation associated with solution of Burgers' equation
\begin{align}\label{eq:impBurg}
\uu(\xx,t)=\uu(\xx-(t-s)\uu(\xx,t),s)\iff \ \frac{D\uu}{Dt}=\frac{\partial \uu}{\partial t}+\frac{\partial \uu}{\partial \xx}\uu=\mb{0}
\end{align}
for $s\leq t$ \cite{evans:2010:pde}. Traditionally, graphics applications have preferred the explicit variant of semi-Lagragian advection whereby grid velocities are updated through the expression
\begin{align}\label{eq:SL}
\uu_\ii^{n+1}=\uu(\xx_\ii-\Delta t \uu_\ii^n,t^n)
\end{align}
where $\xx_\ii$ is the location of grid node $\ii$, $\uu_\ii^n,\uu_\ii^{n+1}$ are velocities at the node at times $t^n$ and $t^{n+1}$ respectively and interpolation over the velocity grid is used to estimate $\uu(\xx_\ii-\Delta t \uu_\ii^n,t^n)$ at non-grid node locations \cite{sawyer:1963:semi,stam:1999:stable}. In contrast, BSL techniques leverage Equation ~\eqref{eq:impBurg} directly 
\begin{align}\label{eq:BSL}
\uu_\ii^{n+1}=\uu(\xx_\ii-\Delta t \uu_\ii^{n+1},t^n)
\end{align}
which requires the solution of an implicit equation for $\uu_\ii^{n+1}$ \cite{robert:1981:stable}. Since our grid interpolation is naturally $C^1$, we show that this can be done very efficiently using a few steps of Newton's method. While this is more expensive than the explicit semi-Lagrangian formulations, we note that each node can still be updated in parallel since the implicit equations for $\uu_\ii^{n+1}$ are decoupled in $\ii$. We show that solution of the implicit Equation~\eqref{eq:BSL}, rather than the traditionally used explicit Equation~\eqref{eq:SL} improves the order of convergence from first to second (in space and time). Notably, this does not require use of multiple time steps for backward/forward estimations of error, as is commonly done \cite{kim:2006:advections,kim:2005:bfecc,selle:2008:unconditionally,xiu:2001:semi,schroeder:2014:vna}. Furthermore, our method allows for larger-than-CFL time steps and is as stable or more so than explicit semi-Lagrangian formulations.\\
\\
\begin{figure}[t]
  \centering
  \begin{subfigure}{.49\columnwidth}
     \includegraphics[draft=\mydraft,trim={0 60px 0 60px},clip,width=\columnwidth]{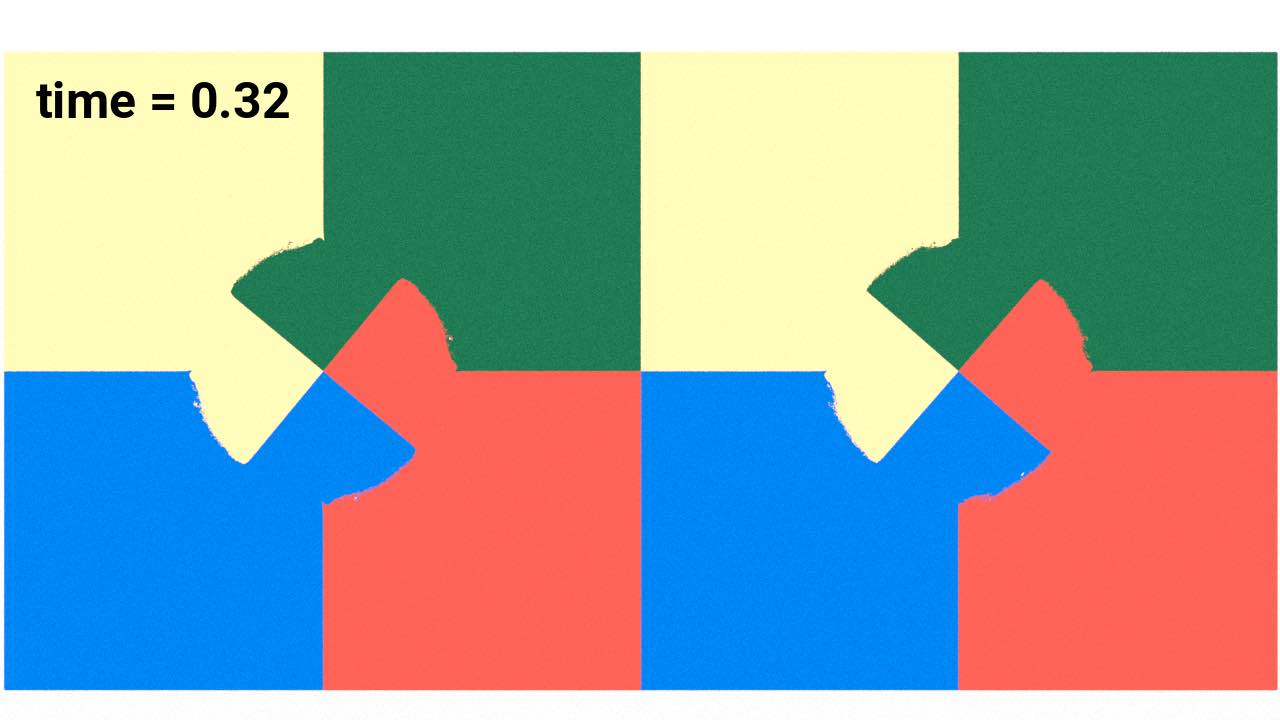}
  \end{subfigure}
  \begin{subfigure}{.49\columnwidth}
     \includegraphics[draft=\mydraft,trim={0 60px 0 60px},clip,width=\columnwidth]{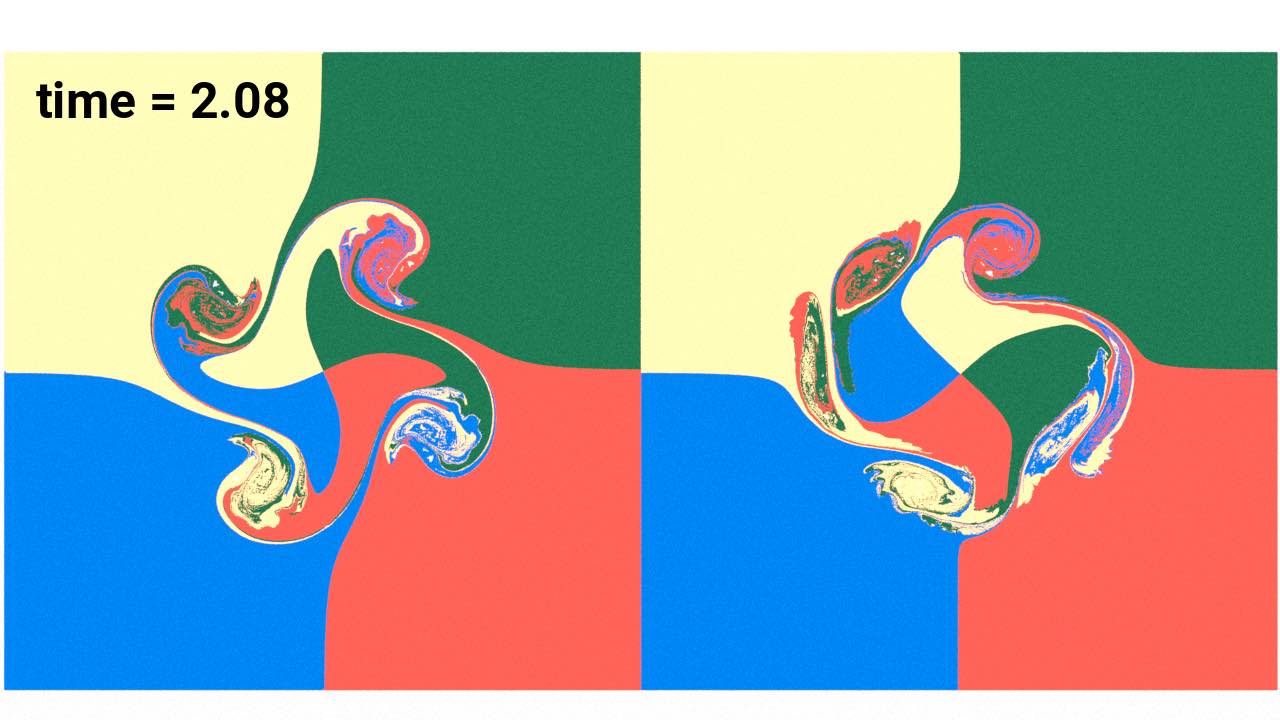}
  \end{subfigure}\\
  \begin{subfigure}{.49\columnwidth}
     \includegraphics[draft=\mydraft,trim={0 60px 0 60px},clip,width=\columnwidth]{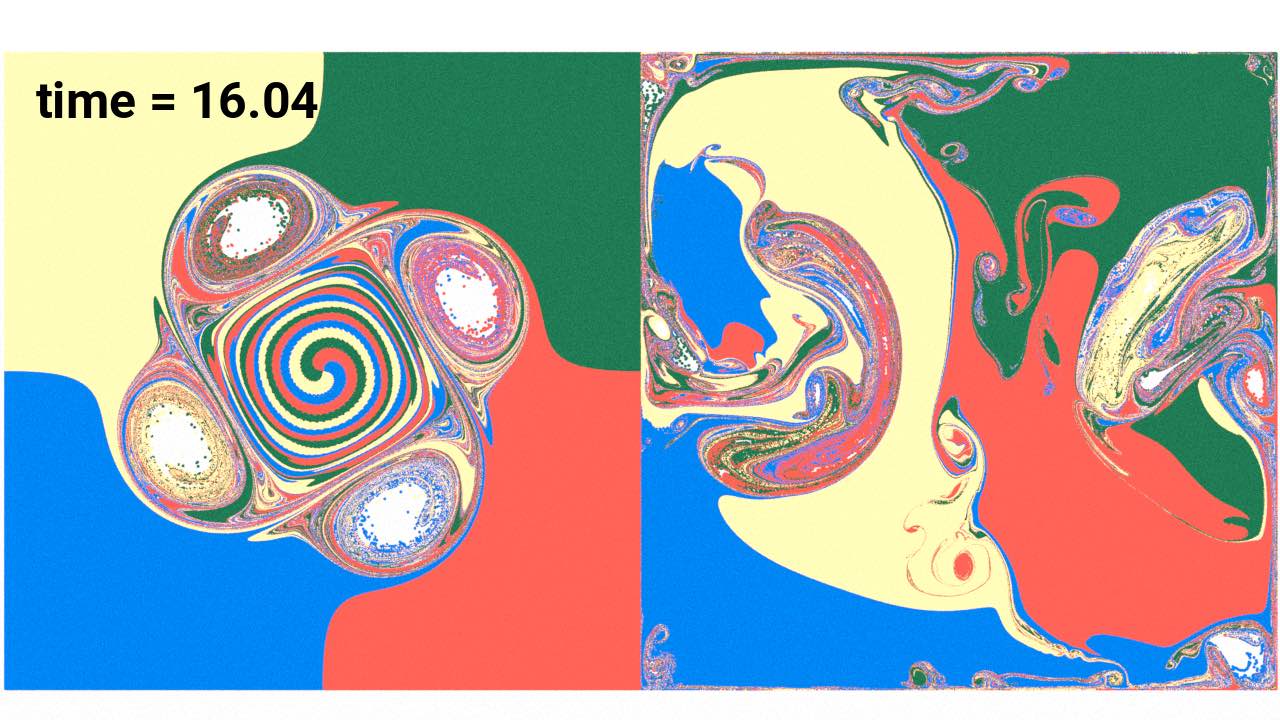}
  \end{subfigure}
  \begin{subfigure}{.49\columnwidth}
     \includegraphics[draft=\mydraft,trim={0 60px 0 60px},clip,width=\columnwidth]{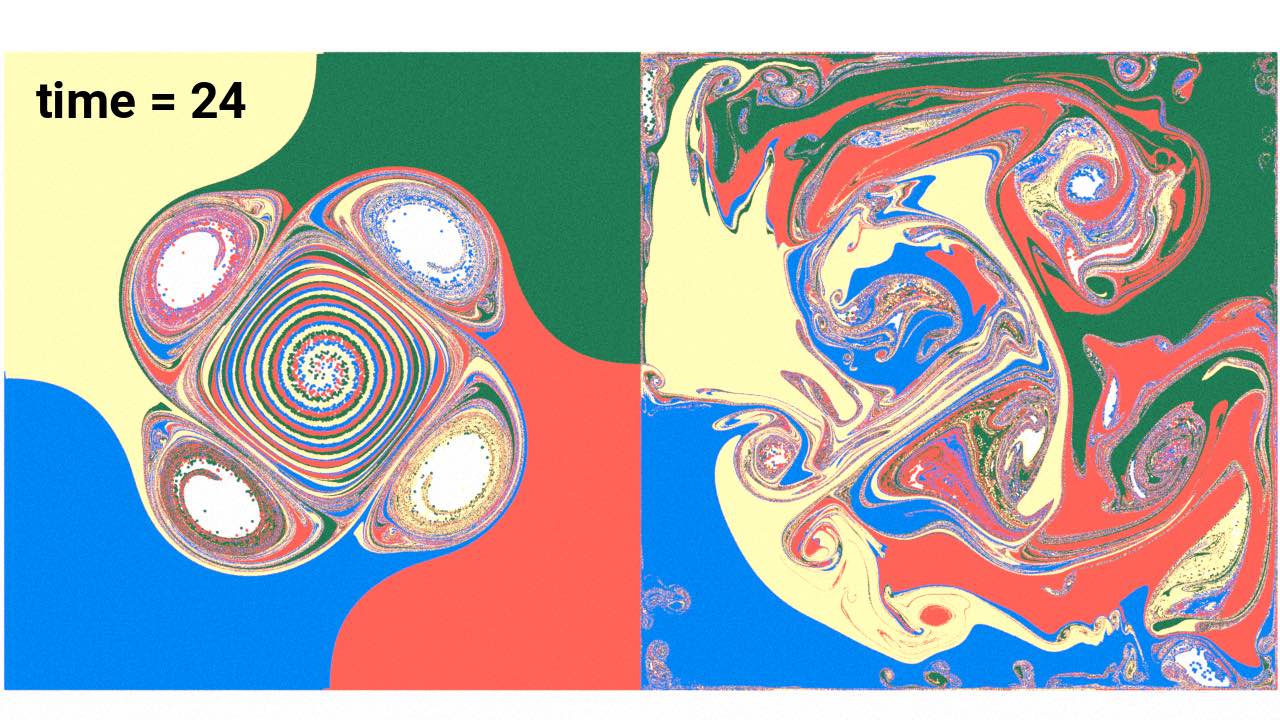}
  \end{subfigure}
\caption{{\textbf{SL vs. BSLQB}}. We compare semi-Lagrangian (left) and BSLQB  (right) in a vorticity-intensive example. BSLQB breaks symmetry and exhibits a more turbulent flow pattern. Note we only use particles for flow visualization and not for PolyPIC advection in this example.}\label{fig:inner_circle}
\end{figure}
Lastly, we develop a hybrid particle/BSLQB advection technique that utilizes PolyPIC \cite{fu:2017:poly} in portions of the domain covered by particles and BSLQB in portions without particles. Our formulation naturally leverages the strengths of both approaches. Dense concentrations of particles can be added to regions of the domain where more detail is desired. Also, if particle coverage becomes too sparse because of turbulent flows, BSLQB can be used in the gaps. We demonstrate the efficacy of this technique with smoke simulation and narrow banding of particles near the fluid surface with water simulations as in \cite{chentanez:2015:coupling,ferstl:2016:narrow,sato:2018:nb}. In this case, level set advection naturally enabled with our BSLQB formulation is preferred in deeper water regions. We summarize our contributions as:
\begin{itemize}
\item A novel cut-cell collocated velocity B-spline mixed FEM method for Chorin \shortcite{chorin:1967:numerical} splitting discretization of the incompressible Euler equations.
\item BSLQB: a novel BSL technique designed for collocated multiquadratic B-spline velocity interpolation that achieves second order accuracy in space and time.
\item A hybrid BSLQB/PolyPIC method for narrow band free-surface flow simulations and concentrated-detail smoke simulations.
\end{itemize}

\begin{figure*}[!ht]
  \centering
  \begin{subfigure}{.33\textwidth}
     \includegraphics[draft=\mydraft,trim={0 0 0 0},clip,width=\columnwidth]{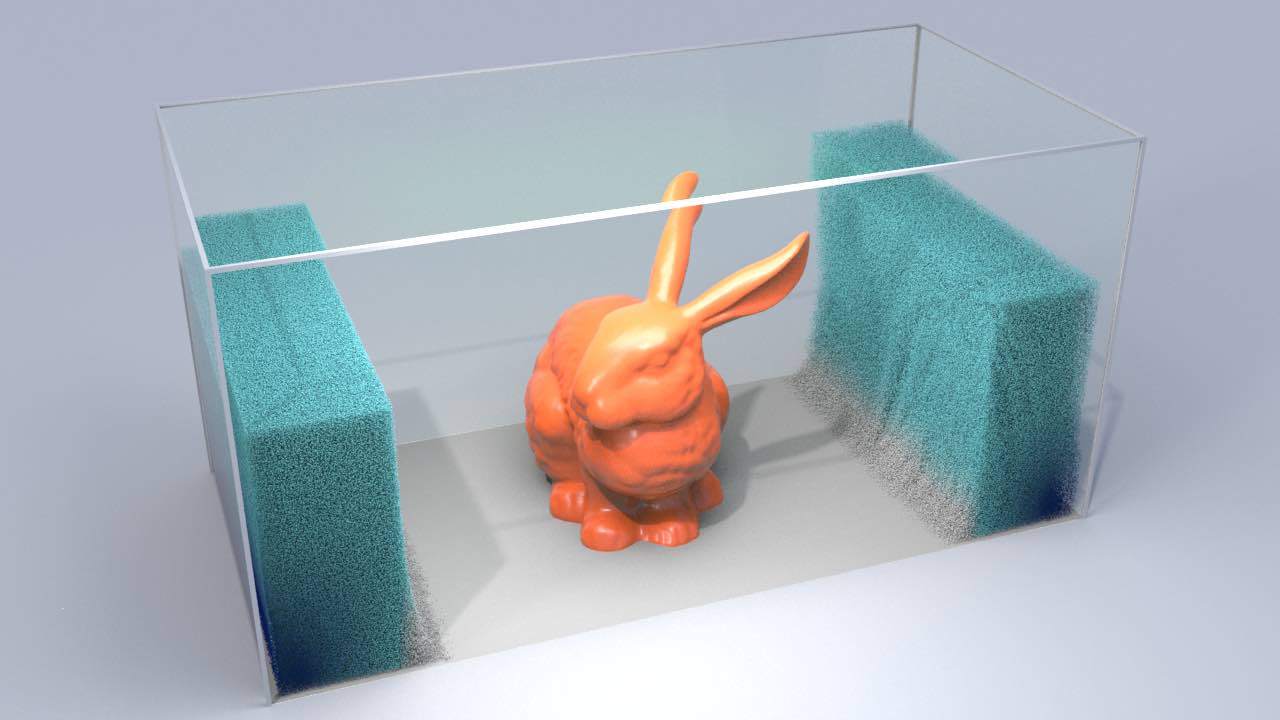}
  \end{subfigure}  
  \begin{subfigure}{.33\textwidth}
     \includegraphics[draft=\mydraft,trim={0 0 0 0},clip,width=\columnwidth]{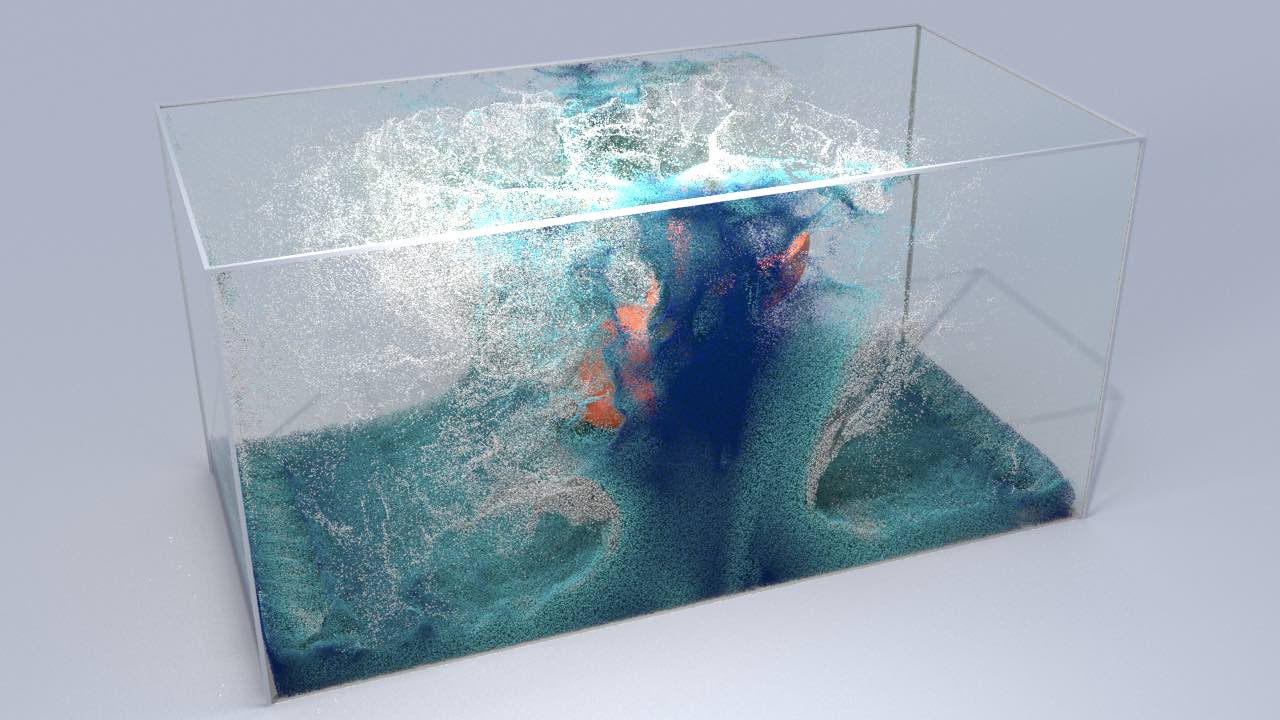}
  \end{subfigure}  
  \begin{subfigure}{.33\textwidth}
     \includegraphics[draft=\mydraft,trim={0 0 0 0},clip,width=\columnwidth]{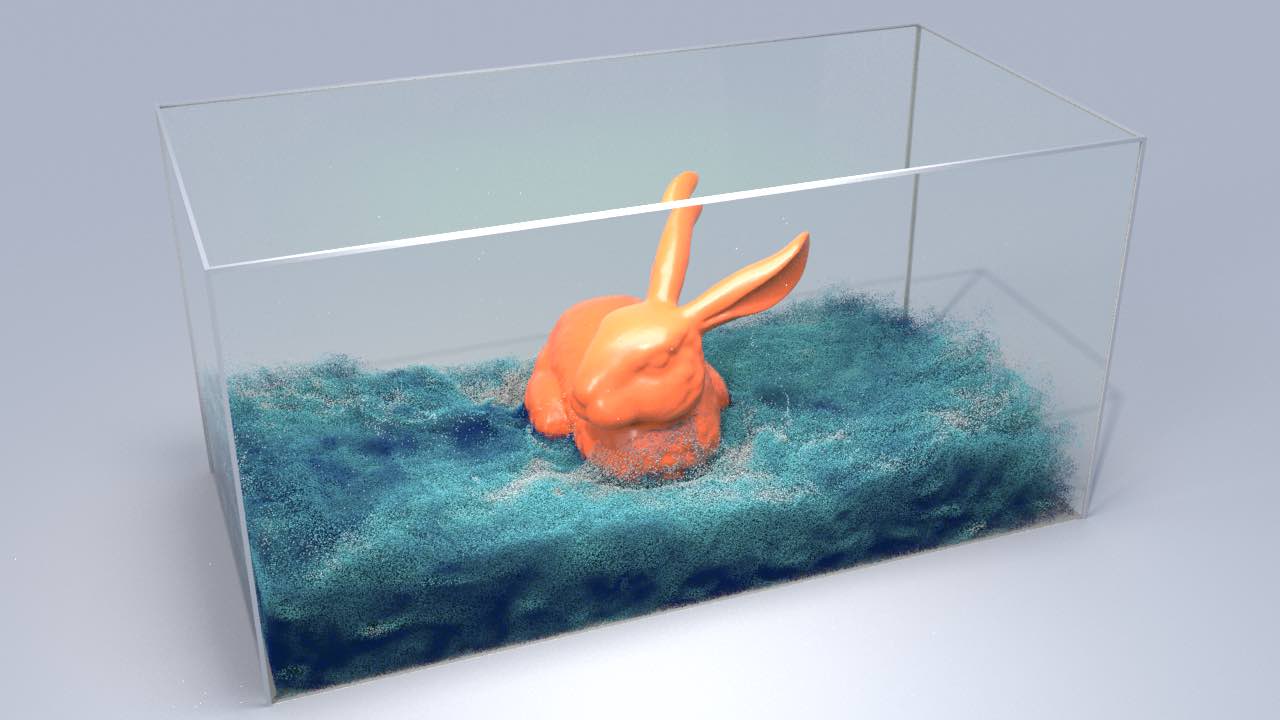}
  \end{subfigure}
\caption{{\textbf{Dam break with bunny}}: Opposing blocks of water collapse in a tank and flow around the irregular domain boundary placed in the middle of the tank.  Particles are colored from slow (blue) to fast (white) speed. }\label{fig:bunny_dambreak_final}
\end{figure*}

\section{Previous Work}

\subsection{Advection}
Stam \shortcite{stam:1999:stable} first demonstrated the efficacy of semi-Lagrangian techniques for graphics applications and they have since become the standard, largely due to the large time steps they engender and their simple interpolatory nature. Many modifications to the original approach of Stam \shortcite{stam:1999:stable} have been developed, often inspired by approaches in the engineering literature. Fedkiw et al. \shortcite{fedkiw:2001:visual} use vorticity confinement \cite{steinhoff:1994:modification} to counterbalance vorticity lost to dissipation and cubic grid interpolation. Kim et al. \shortcite{kim:2006:advections,kim:2005:bfecc} and Selle et al. \cite{selle:2008:unconditionally} combine forward and backward semi-Lagrangian steps to estimate and remove dissipative errors. Constrained Interpolation Profile \cite{kim:2008:semi,yabe:2001:multiphase-analysis,song:2009:derivative-particles} techniques additionally advect function derivatives to reduce dissipation. Molemaker et al. \shortcite{molemaker:2008:low} use the QUICK technique of Leonhard \shortcite{leonard:1979:stable} which is essentially upwinding with quadratic interpolation and Adams-Bashforth temporal discretization, although this does not have the favorable stability properties of semi-Lagrangian. Backward Difference Formula techniques are useful because they use an implicit multistep formulation for higher-order semi-Lagrangian advection yet still only require one projection per time step \cite{xiu:2001:semi,schroeder:2014:vna}.\\
\\
The main idea in semi-Lagrangian techniques is to interpolate data from a characteristic point. This idea goes back to the Courant-Issaacson-Rees \shortcite{courant:1952:solution} method. However, as noted in \cite{fedkiw:2001:visual} semi-Lagrangian advection is very popular in atmospheric science simulation and the variants used in graphics that account for characteristics traveling beyond the local cell in one time step go back to Sawyer \shortcite{sawyer:1963:semi}. The first BSL approach utilizing Equation~\eqref{eq:BSL} was done by Robert \shortcite{robert:1981:stable} in which they use fixed point iteration to solve the nonlinear equation. They fit a bicubic function to their data over $4\times4$ grid patches, then use that function in the fixed point iteration. If the upwind point leaves the grid, they clamp it to the boundary of the $4\times4$ patch. This clamping will degrade accuracy for larger time steps. In this case, more general interpolation is typically used (see \cite{staniforth:1991:semi,falcone:1998:convergence} for useful reviews). Pudykiewicz and Staniforth \shortcite{pudykiewicz:1984:some} investigate the effects of BSL versus explicit semi-Lagrangian. Specifically, they compare Bates and McDonald \shortcite{bates:1982:multiply} (explicit) versus Robert \shortcite{robert:1981:stable} (BSL). They show that keeping all things equal, the choice of Equation~\eqref{eq:SL} (explicit) instead of Equation~\eqref{eq:BSL} (BSL) leads to more dissipation and mass loss. This is consistent with our observations with BSLQB.\\
\\
Interestingly, multiquadratic B-splines have not been adopted by the semi-Lagrangian community, despite their natural regularity. Hermite splines, multicubic splines and even Lagrange polynomials are commonly used \cite{staniforth:1991:semi}. Preference for Hermite splines and Lagrange polynomials is likely due to their local nature (they do not require solution of a global system for coefficients) and preference for multicubic splines (over multi-quadratic) is possibly due to the requirement of odd degree for natural splines (odd degree splines behave like low pass filters and tend to be smoother than even degree splines \cite{cheng:2001:quadratic,cheney:2012:numerical}). Cubic splines are considered to be more accurate than Hernite splines and Lagrange interpolation \cite{staniforth:1991:semi,makar:1996:basis}. Interestingly, Riish{\o}jgaard et al. \shortcite{riishojgaard:1998:use} found that cubic spline interpolation gave rise to a noisier solution than cubic Lagrange interpolation with a technique analogous to that of Makar and Karpik \shortcite{makar:1996:basis}. However, they also note that addition of a selective scale diffusion term helps reduce noise associated with cubic splines. Wang and Layton \shortcite{wang:2010:new} use linear B-splines with BSL but only consider one space dimension which makes Equation~\eqref{eq:BSL} linear and easily solvable.\\
\\
Dissipation with explicit semi-Lagrangian advection is so severe that many graphics researchers have resorted to alternative methods to avoid it. Mullen et al. \shortcite{mullen:2009:energy} develop energy preserving integration to prevent the need for correcting dissipative behavior. Some authors \cite{qu:2019:mcm,tessendorf:2011:MCM,sato:2017:long,sato:2018:spatially} resolve the flow map characteristics for periods longer than a single time step (as opposed to one step with semi-Lagrangian) to reduce dissipation. Hybrid Lagrange/Eulerian techniques like PIC (and related approaches) \cite{bridson:2008:fluid-simulation,jiang:2015:apic,fu:2017:poly,zhu:2005:sand-fluid} explicitly track motion of particles in the fluid, which is nearly dissipation-free, but can suffer from distortion in particle sampling quality. Vorticity formulations are also typically less dissipative, but can have issues with boundary conditions enforcement \cite{selle:2005:vortex,angelidis:2005:simulation,chern:2016:schrodinger,elcott:2007:stable,park:2005:vortex,weissmann:2010:filament}. Zehnder et al., Zhang et al. and Mullen et al. \shortcite{mullen:2009:energy,zehnder:2018:advection,narain:2019:ref,zhang:2015:restoring} have noted that the Chorin projection itself causes dissipation. Zhang et al. \shortcite{zhang:2015:restoring} reduced artificial dissipation caused by the projection step by estimating lost vorticity and adding it back into the fluid. Zehnder et al. \shortcite{zehnder:2018:advection,narain:2019:ref} propose a simple, but very effective modification to the splitting scheme that is similar to midpoint rule integration to reduce the projection error.

\subsection{Pressure projection}

Graphics techniques utilizing pressure projection typically use voxelized MAC grids with boundary conditions enforced at cell centers and faces, however many methods improve this by taking into account sub-cell geometric detail. Enright et al. \shortcite{enright:2003:using} showed that enforcing the pressure free surface boundary condition at MAC grid edge crossings (rather than at cell centers) dramatically improved the look of water surface waves and ripples. Batty, Bridson and colleagues developed variational weighted finite difference approaches to enforce velocity boundary conditions with MAC grids on edge crossings and improved pressure boundary conditions at the free surface in the case of viscous stress \cite{batty:2007:solid-fluid,batty:2008:buckling,larionov:2017:stokes}. XFEM \cite{belytschko:2009:review,koschier:2017:xfem} and virtual node (VNA) \cite{schroeder:2014:vna} techniques also use cut cell geometry with variational techniques. Schroeder et al. \shortcite{schroeder:2014:vna} use cut cells with MAC grids, but their technique is limited to moderate Reynolds numbers. \\
\\
\begin{figure}[!t]
  \centering
  \begin{subfigure}{0.24\columnwidth}
     \includegraphics[draft=\mydraft,trim={300px 0 300px 0},clip,width=\columnwidth]{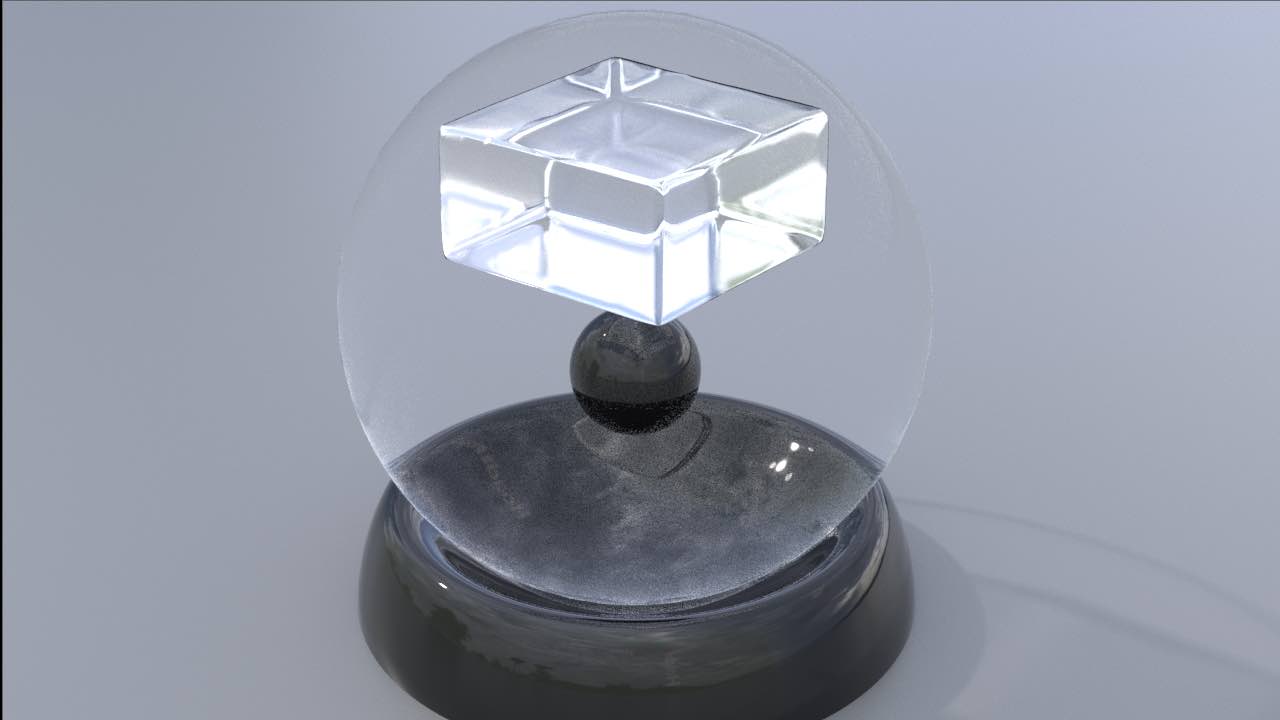}
  \end{subfigure}
  \begin{subfigure}{0.24\columnwidth}
     \includegraphics[draft=\mydraft,trim={300px 0 300px 0},clip,width=\columnwidth]{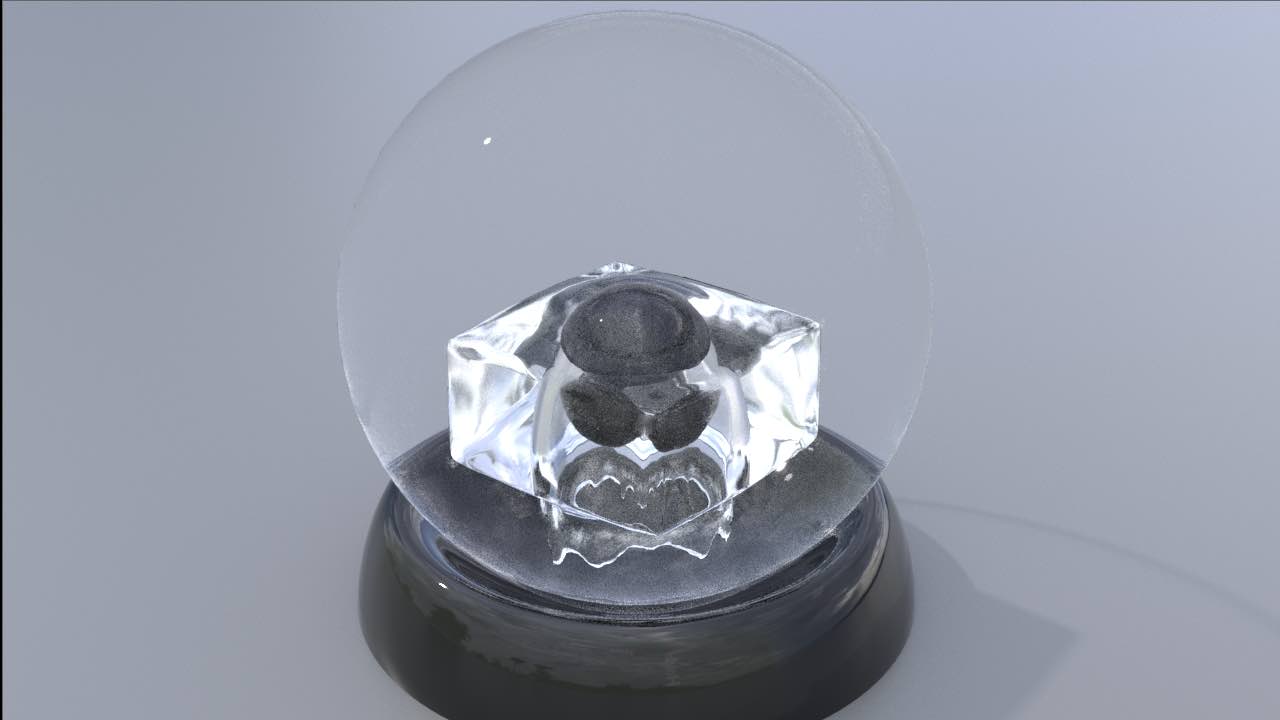}
  \end{subfigure} 
  \begin{subfigure}{0.24\columnwidth}
     \includegraphics[draft=\mydraft,trim={300px 0 300px 0},clip,width=\columnwidth]{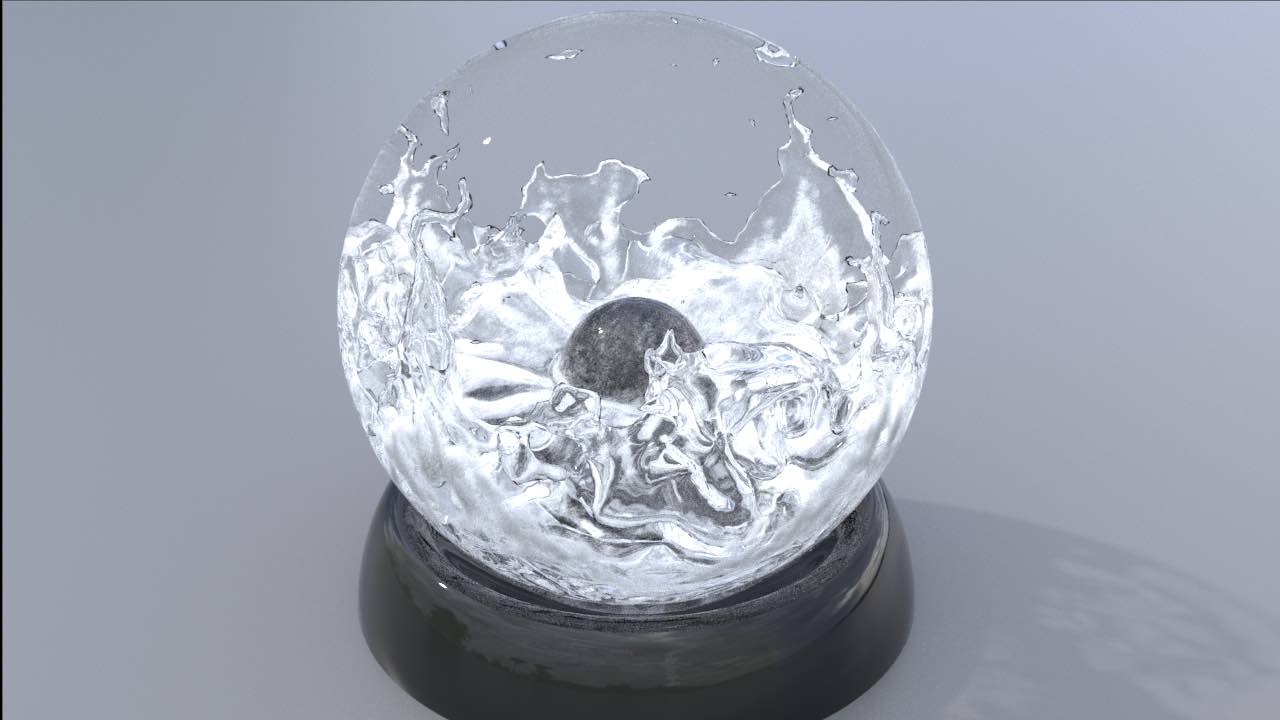}
  \end{subfigure}
  \begin{subfigure}{0.24\columnwidth}
     \includegraphics[draft=\mydraft,trim={300px 0 300px 0},clip,width=\columnwidth]{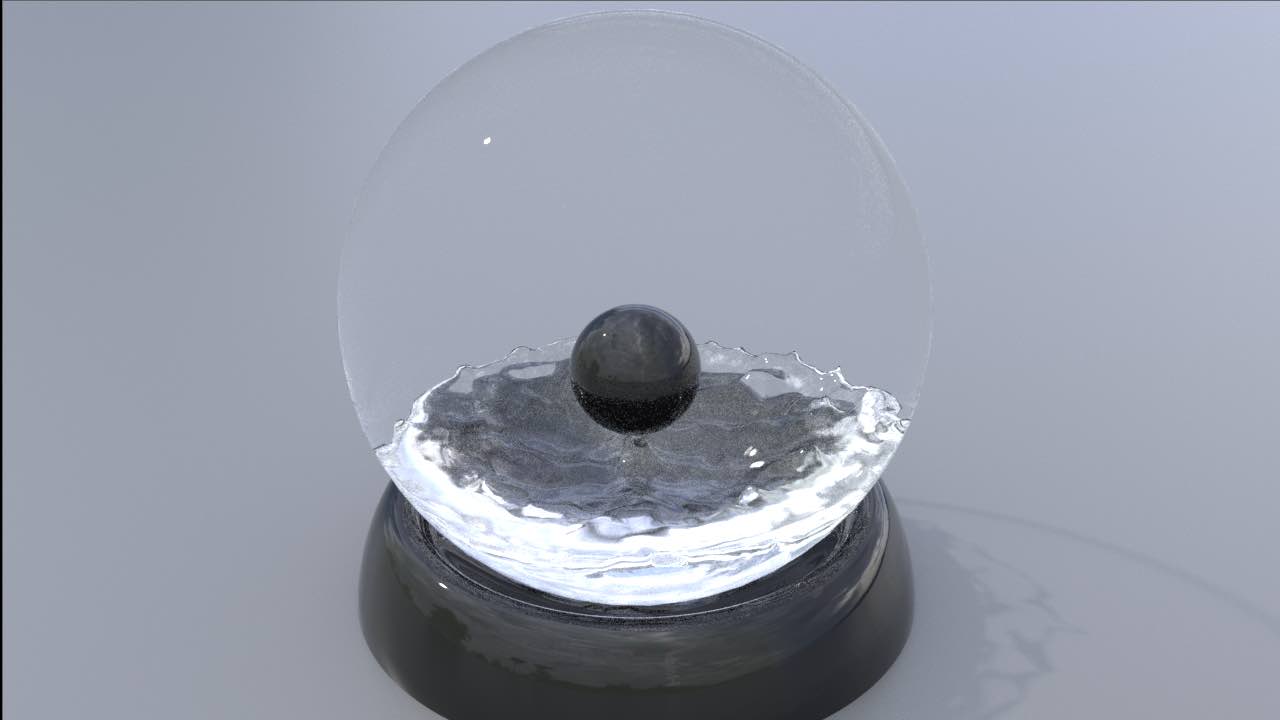}
  \end{subfigure}
\caption{{\textbf{Water in a globe}}. A block of water splashes and naturally slides along cut cell boundaries in an irregular domain interior to one large sphere and exterior to one small sphere.}\label{fig:snowglobe}
\end{figure}
There is a vast literature on enforcing incompressibility in the FEM community \cite{hughes:2000:book}. Our approach is most similar to the B-spline Taylor-Hood element of Bressan \cite{bressan:2010:isogeometric}. Adoption of B-spline interpolation in FEM is part of the isogeometric movement \cite{hughes:2005:isogeometric,ruberg:2012:subdivision}. Originally motivated by the desire to streamline the transition from computer-aided design (CAD) to FEM simulation, isogeometric analysis explores the use of CAD-based interpolation (e.g. B-splines and nonuniform rational B-splines (NURBS)) with FEM methodologies. Hughes et al. \shortcite{hughes:2005:isogeometric} show that in addition to simplifying the transition from CAD to simulation, the higher regularity and spectral-like properties exhibited by these splines makes them more accurate than traditionally used interpolation. We enforce Dirichlet boundary conditions weakly as in XFEM and VNA approaches \cite{belytschko:2009:review,koschier:2017:xfem,schroeder:2014:vna}. Bazilevs et al. \shortcite{bazilevs:2007:weak} show that weak Dirichlet enforcement with isogeometric analysis can be more accurate than strong enforcement.\\
\\
Graphics applications are typically concerned with turbulent, high-Reynolds numbers flows. Interestingly, B-splines have proven effective for these flows by researchers in the Large Eddy Simulation (LES) community \cite{kim:1998:mixed,kravchenko:1999:bspline}. Kravchenko et al. \shortcite{kravchenko:1999:bspline} use a variational weighted residuals approach with B-splines for turbulent LES and show that the increased regularity significantly reduces computational costs. Boatela et al. \shortcite{botella:2002:collocation} use a similar approach, but apply a collocation technique where the strong form of the div-grad formulation of incompressibility is enforced point wise. They show that their B-spline approach attains optimal order of accuracy with accurate resolution of quadratic flow invariants. Boatela et al. \shortcite{botella:2002:collocation} also introduce a notion of sparse approximation to the inverse mass matrix to avoid dense systems of equations in the pressure solve.

\begin{figure}[t]
  \centering
  \begin{subfigure}{0.5\columnwidth}
     \includegraphics[draft=\mydraft,trim={300px 0 300px 0},clip,width=\columnwidth]{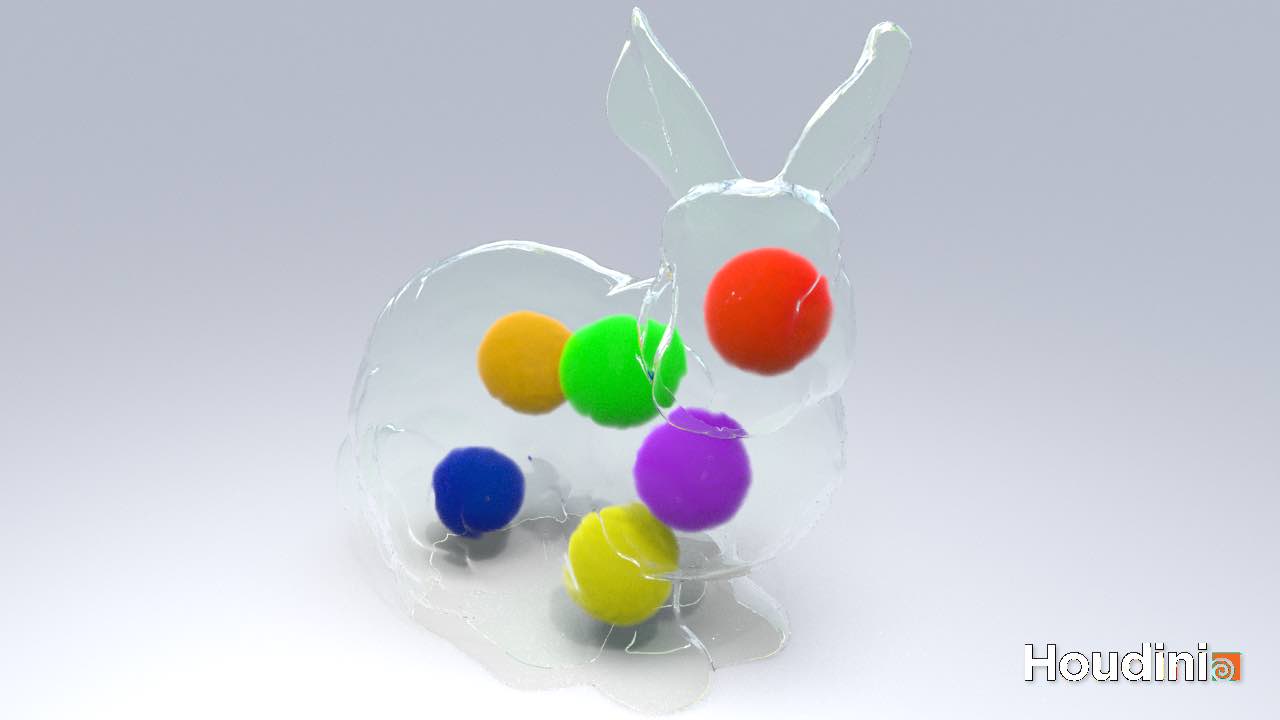}
  \end{subfigure}
  \begin{subfigure}{0.5\columnwidth}
     \includegraphics[draft=\mydraft,trim={300px 0 300px 0},clip,width=\columnwidth]{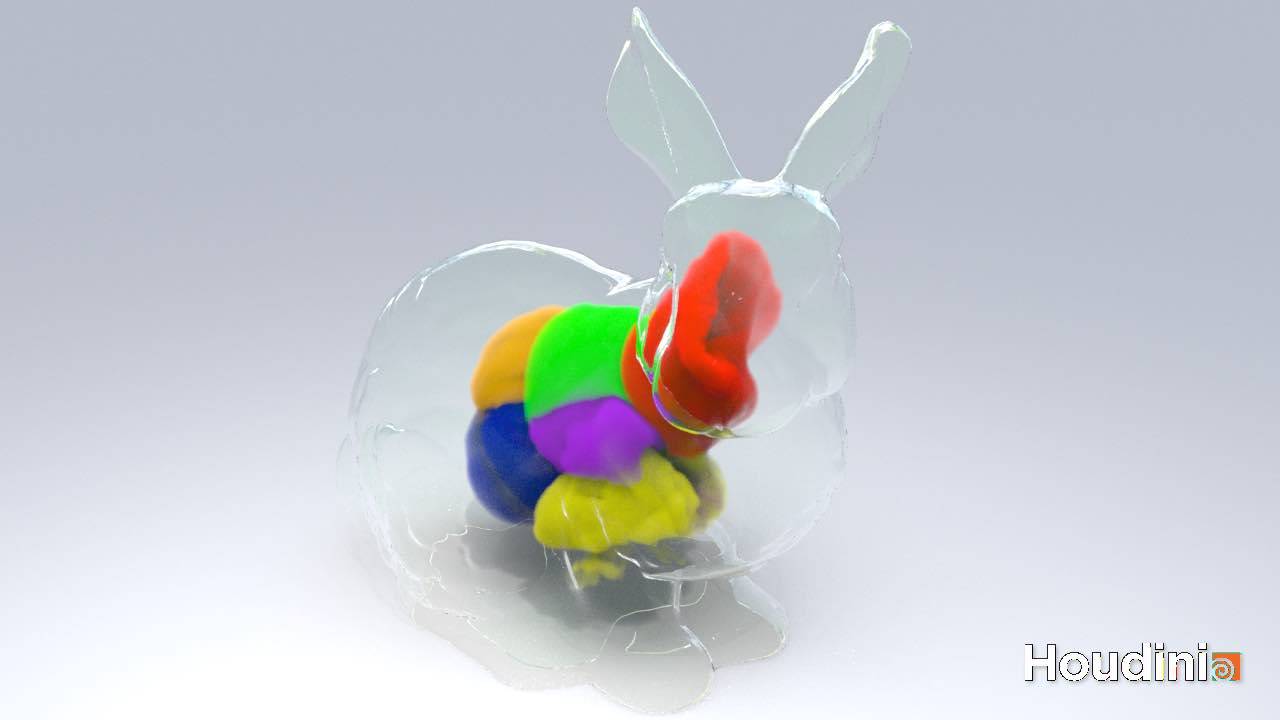}
  \end{subfigure} \\[-1ex] 
  \begin{subfigure}{0.5\columnwidth}
     \includegraphics[draft=\mydraft,trim={300px 0 300px 0},clip,width=\columnwidth]{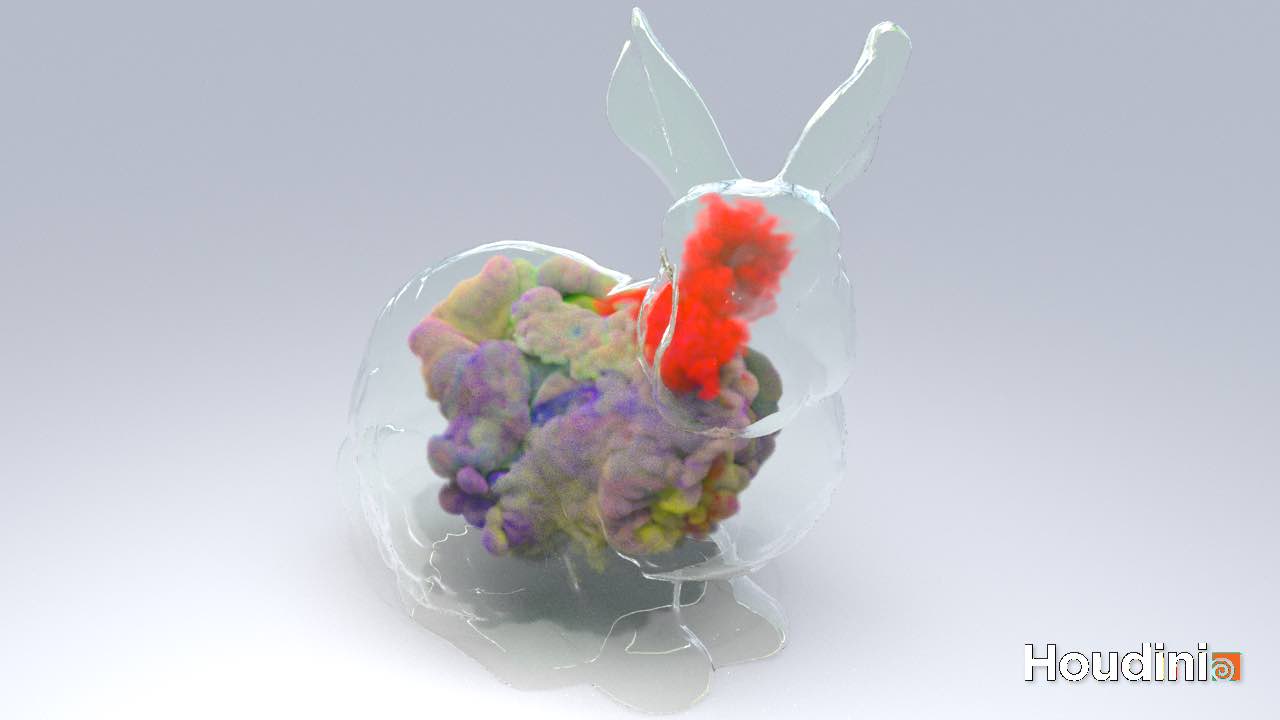}
  \end{subfigure}
  \begin{subfigure}{0.5\columnwidth}
     \includegraphics[draft=\mydraft,trim={300px 0 300px 0},clip,width=\columnwidth]{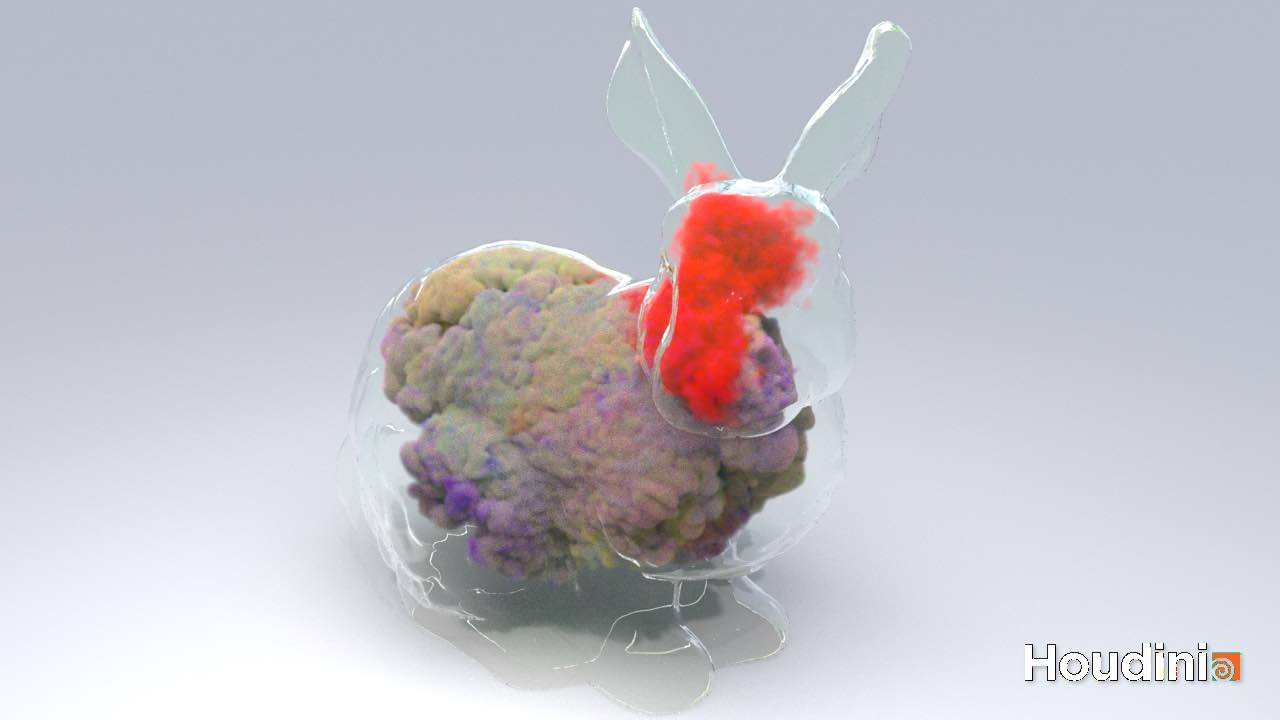}
  \end{subfigure}
\caption{{\textbf{Smoke in an irregular domain}}. Multicolored spheres of smoke with non-zero initial velocity conditions flow and collide inside the Stanford bunny. Zero normal velocity is enforced with our cut cell formulation.}\label{fig:bunnysmoke}
\end{figure}

\section{Governing Equations and Operator Splitting}
\begin{figure}[h]
\centering
\begin{tabular}{cc}
\subf{\includegraphics[draft=\mydraft,width=.45\columnwidth]{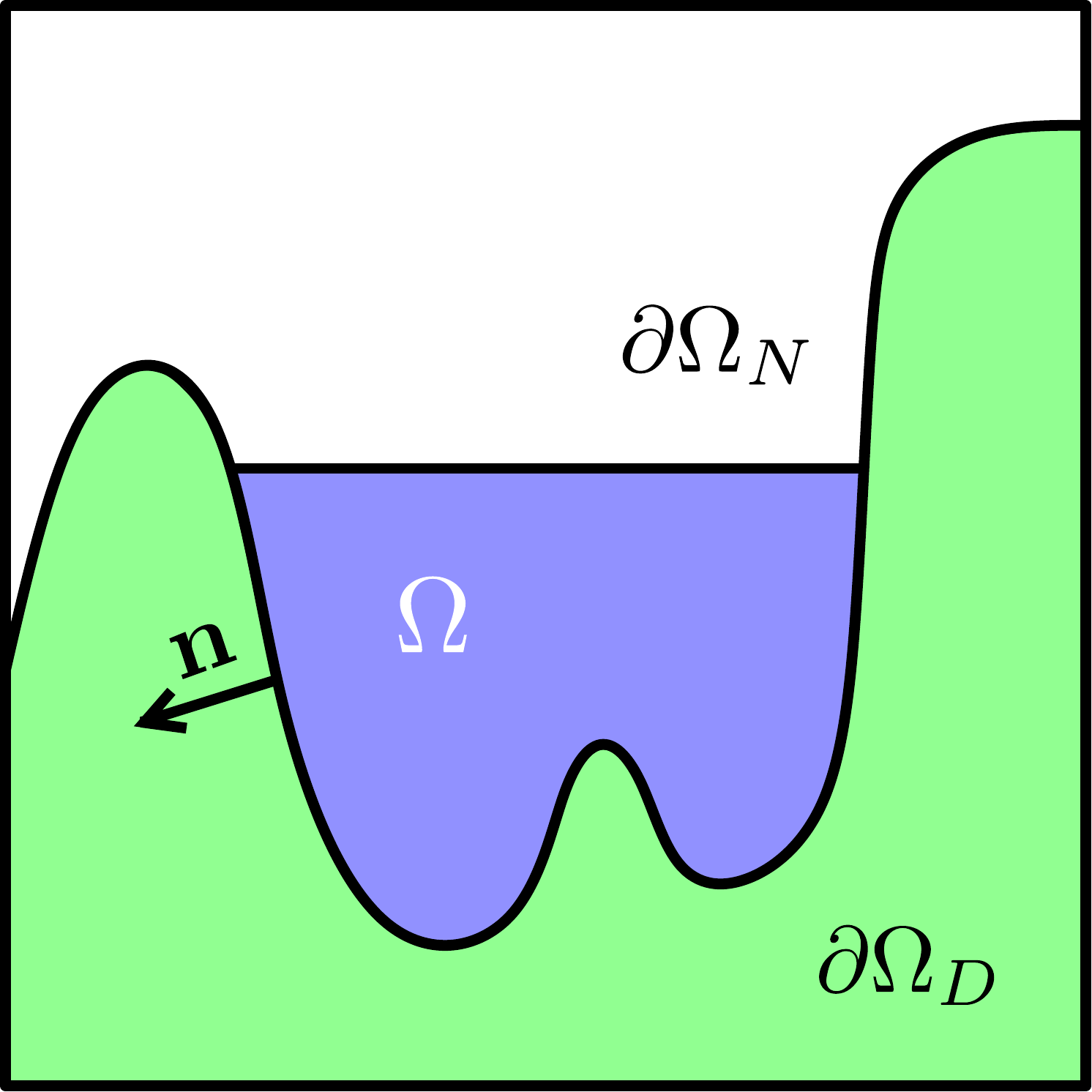}}
&
\subf{\includegraphics[draft=\mydraft,width=.45\columnwidth]{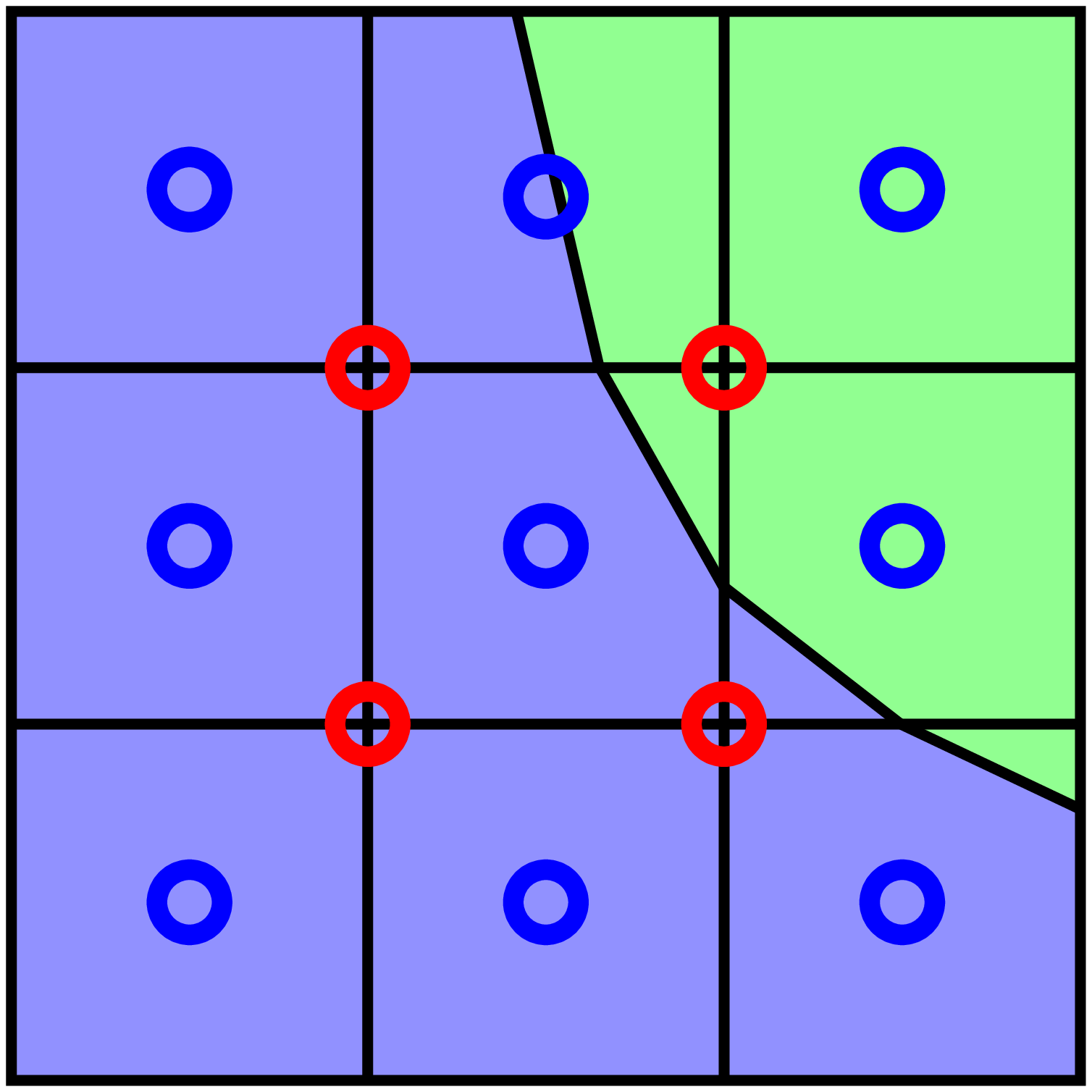}}
\\
\end{tabular}
\caption{{\textbf{Flow domain and grid}}. {\textbf{Left}}: we use $\Omega$ to denote the fluid domain, with $\partial\Omega_D$ used to indicate the portion of the fluid domain subject to velocity boundary conditions and $\partial\Omega_N$ to indicate the free-surface portion of the boundary with pressure condition $p=0$. {\textbf{Right}}: We use multiquadratic interpolation for velocity ($\bar{\uu}_\ii$ at cell centers, blue) and multilinear for pressure ($p_\cc$ at nodes, red). The fluid domain is defined with sub-grid-cell accuracy.}\label{fig:dAndg}
\end{figure}

\begin{figure*}[ht!]
\includegraphics[draft=\mydraft,width=\textwidth]{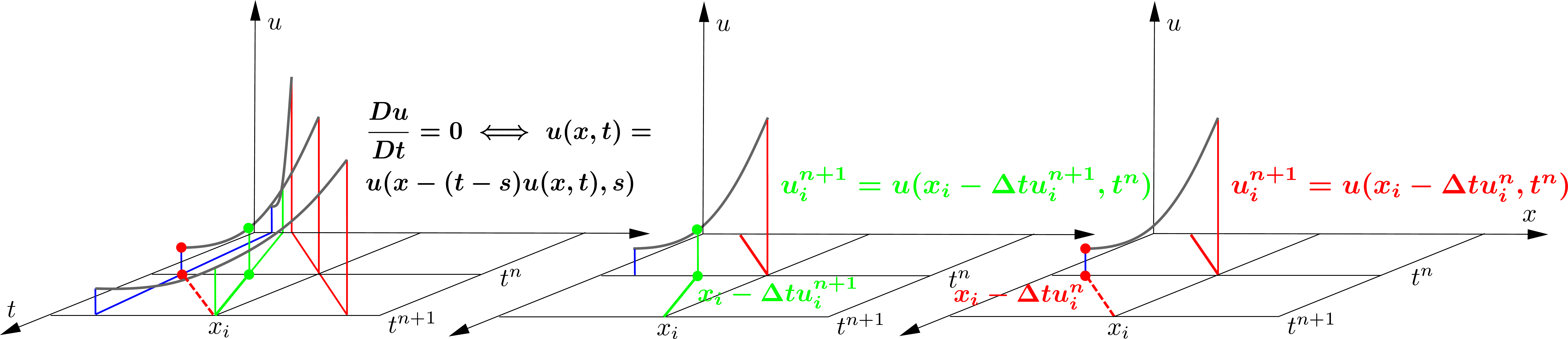}
\caption{{\textbf{BSL versus SL}}. We illustrate the difference between explicit semi-Lagrangian and BSL in 1D. {\textbf{Left}}: The exact solution of Burgers' equation has straight line characteristics shown in blue, green and red on which velocity (plotted above the plane in gray) is constant. {\textbf{Center}}: BSL (green) uses Newton's method to solve for the exact characteristic going through $x_i$ at time $t^{n+1}$ to determine $u_i^{n+1}$. {\textbf{Right}}: explicit semi-Lagrangian (red) uses a stale, time $t^n$ approximation of the characteristic which over shoots, resulting in an underestimate of the velocity and energy loss.}\label{fig:burgers}
\end{figure*}

We solve the incompressible Euler equations that describe the evolution of a fluid in terms of its mass density $\rho$, velocity $\uu$, pressure $p$ and gravitational constant $\gg$ as
\begin{align}
\rho\frac{D\uu}{Dt} &=\rho\left(\frac{\partial \uu}{\partial t} + \frac{\partial \uu}{\partial \xx}\uu\right)=-\nabla p + \rho\gg, \ \xx\in\Omega \label{eq:mom_cont}\\
\nabla\cdot\uu &= 0, \ \xx\in\Omega \label{eq:div_cont} \\
\uu\cdot\nn&=a, \ \xx\in\partial \Omega_D \label{eq:bcv_cont}\\
p&=0, \ \xx\in\partial \Omega_N \label{eq:bcp_cont}
\end{align}
where Equation~\eqref{eq:mom_cont} is balance of linear momentum, Equation~\eqref{eq:div_cont} is the incompressibility constraint, Equation~\eqref{eq:bcv_cont} is the boundary condition for the normal component of the velocity and Equation~\eqref{eq:bcp_cont} is the free surface boundary condition. We use $\Omega$ to denote the region occupied by the fluid, $\partial \Omega_D$ to denote the portion of the boundary of the fluid domain on which velocity is prescribed to be $a$ (which may vary over the boundary) and $\partial \Omega_N$ is the surface of the water where the pressure is zero (see Figure~\ref{fig:dAndg}).\\
\\
In a Chorin \shortcite{chorin:1967:numerical} operator splitting of the advective and pressure terms, velocity is first updated to an intermediate field $\ww$ under the convective $\rho\frac{D\uu}{Dt}=\mb{0}$, followed by an update from the pressure and gravitational body forcing under $\rho\frac{\partial \uu}{\partial t}=-\nabla p + \rho\gg$ where the pressure is determined to enforce $\nabla\cdot\uu = 0$. Dividing by the mass density, the convective step is seen to be an update under Burgers' equation~\eqref{eq:impBurg}. Burgers' equation governs temporally constant Lagrangian velocity (zero Lagrangian acceleration). The characteristic curves for flows of this type are straight lines (since the Lagrangian acceleration is zero), on which the velocity is constant (see Figure~\ref{fig:burgers}). This gives rise to the implicit relation $\uu(\xx,t)=\uu(\xx-(t-s)\uu(\xx,t),s)$ for $s\leq t$. Intuitively, if we want to know the velocity $\uu(\xx,t)$ at point $\xx$ at time $t$, we look back along the characteristic passing through $\xx$ at time $t$ to any previous time $s$; however, the characteristic is the straight line defined by the velocity $\uu(\xx,t)$ that we want to know. Hence we take an implicit approach to the solution of this equation, which when combined with the operator splitting amounts to
\begin{align}
\frac{\ww-\tilde{\uu}^n}{\Delta t} &=\mb{0} \label{eq:split_a} \\
\rho\frac{\uu^{n+1}-\ww}{\Delta t} &=-\nabla p^{n+1} + \rho\gg \label{eq:split_p}\\
\nabla\cdot\uu^{n+1} &= \mb{0} \label{eq:split_div}
\end{align}
where we use the notation $\uu^{n+\alpha}(\xx)=\uu(\xx,t^{n+\alpha})$, $\alpha=0,1$ to denote the time $t^{n+\alpha}$ velocities. Furthermore, the intermediate velocity $\ww$ is related to $\tilde{\uu}^n$ through $\tilde{\uu}^n(\xx)=\uu(\xx-\Delta t \ww(\xx),t^n)$.

\section{Spatial Discretization}
We discretize in space by first representing velocity and pressure in terms of mulitquadratic and multilinear B-splines for velocity and pressure respectively. We use a regular grid with spacing $\Delta x$ and define pressure degrees of freedom at grid vertices and velocity degrees of freedom at grid cell centers as in \cite{ando:2013:surfacing} (see Figure~\ref{fig:dAndg}). This efficiently aligns the support of the multiquadratic and multilinear interpolating functions which naturally allows for a grid-cell-wise definition of the flow domain (see Figure~\ref{fig:fsdnb}). We use $N_\ii(\xx)$ to represent the multiquadratic B-spline basis function associated with velocity degree of freedom $\bar{\uu}_\ii$ at grid cell center $\xx_\ii$ and $\chi_\cc(\xx)$ for the multilinear basis function associated with pressure $p_\cc$ at grid node $\xx_\cc$. These are defined as
\begin{align}
N_\ii(\xx)&=\prod_{\alpha}\hat{N}(\frac{x_\alpha-x_{\alpha \ii}}{\Delta x}), \ \chi_\cc(\xx)=\prod_{\alpha}\hat{\chi}(\frac{x_\alpha-x_{\alpha \cc}}{\Delta x})\\
\hat{N}(\eta)&=\left\{\begin{array}{lcc}
\frac{\left(\eta +\frac{3}{2}\right)^2}{2},&\eta\in(-\frac{3}{2},-\frac{1}{2})\\
-\eta^2+\frac{3}{4},&\eta\in[-\frac{1}{2},\frac{1}{2}]\\
\frac{\left(\eta -\frac{3}{2}\right)^2}{2},&\eta\in(\frac{1}{2},\frac{3}{2})\\
0,&\textrm{otherwise}
\end{array}\right. \ 
\hat{\chi}(\nu)=\left\{\begin{array}{lcc}
1+\nu,&\nu\in(-1,0)\\
1-\nu,&\nu\in[0,1)\\
0,&\textrm{otherwise}
\end{array}\right.
\end{align}
where we use Greek indices $\alpha$ to indicate components of the vectors $\xx$, $\xx_\ii$ and $\xx_\cc$. With this convention we interpolate to define velocity and pressure fields 
\begin{align}\label{eq:interp}
\uu(\xx)=\sum_\ii \bar{\uu}_\ii N_\ii(\xx), \ p(\xx)=\sum_\cc p_\cc \chi_\cc(\xx).
\end{align}
We use the notation $\bar{\uu}_\ii$ to distinguish it from the velocity at the grid node $\uu(\xx_\ii)=\sum_\jj \bar{\uu}_\jj N_\jj(\xx_\ii)$ since the multiquadratic B-splines are not interpolatory and these will in general be different. Note that multilinear interpolation is interpolatory and $p_\cc=\sum_\dd p_\dd \chi_\dd(\xx_\cc)$.
\subsection{BSLQB Advection}\label{sec:bslqb}
With this interpolation choice, we first solve for intermediate grid node velocity values $\ww(\xx_\ii)$ from Equation~\eqref{eq:split_a} as
\begin{align}\label{eq:ad_disc}
\ww(\xx_\ii)=\sum_\jj \bar{\uu}^n_\jj N_\jj\left(\xx_\ii-\Delta t \ww(\xx_\ii)\right).
\end{align}
We can solve this equation using Newton's method since the multiquadratic B-splines are $C^1$. We use $\ww^k_\ii$ to denote the $k^\textrm{th}$ Newton approximation to $\ww(\xx_\ii)$. Explicit semi-Lagrangian is used as an initial guess with $\ww^0_\ii=\sum_\jj \bar{\uu}^n_\jj N_\jj\left(\xx_\ii-\Delta t \sum_\ll \bar{\uu}^n_\ll N_\ll(\xx_\ii)\right)$ and then we update iteratively via $\ww^k_\ii\mathrel{+}=\boldsymbol\delta\uu^k$ with Newton increment $\boldsymbol\delta\uu^k$ satisfying
\begin{align*}
\boldsymbol\delta\uu^k&=\left(\II+\Delta t \frac{\partial \uu^n}{\partial \xx}\left(\xx_\ii-\Delta t \ww^k_\ii\right)\right)^{-1}\left(\sum_\jj \bar{\uu}^n_\jj N_\jj\left(\xx_\ii-\Delta t \ww^k_\ii\right)-\ww^k_\ii\right)
\end{align*}
where $\frac{\partial \uu^n}{\partial \xx}\left(\xx_\ii-\Delta t \ww^k_\ii\right)=\sum_\jj \bar{\uu}^n_\jj\frac{\partial N_\jj}{\partial \xx}\left(\xx_\ii-\Delta t \ww^k_\ii\right)$. It is generally observed \cite{kuo:1990:semi,pudykiewicz:1984:some} that with BSL approaches of this type, this iteration will converge as long as $\II+\Delta t \sum_\jj \bar{\uu}^n_\jj \frac{\partial N_\jj}{\partial \xx}\left(\xx_\ii-\Delta t \ww^k_\ii\right)$ is non-singular. We note that this condition holds as long as no shocks form under Burgers' equation \cite{evans:2010:pde} (forward from time $t^n$). This is a safe assumption since we are modeling incompressible flow with which shock formation does not occur, but it may be a problem for compressible flows. In practice, this iteration converges in 3 or 4 iterations, even with CFL numbers larger than 4 (see Section~\ref{sec:ex_hybrid}). When it does fail (which occurs less than one percent of the time in the examples we run), it is usually for points near the boundary with characteristics that leave the domain (since we cannot estimate $\frac{\partial \uu^n}{\partial \xx}$ using grid interpolation if the upwind estimate leaves the grid). In this case we use explicit semi-Lagrangian and interpolate from the boundary conditions if the characteristic point is off the domain.\\
\\
Once we have obtained the grid node values of the intermediate velocity $\ww(\xx_\ii)$, we must determine interpolation coefficients $\bar{\ww}_\jj$ such that $\ww(\xx_\ii)=\sum_\jj \bar{\ww}_\jj N_\jj(\xx_\ii)$. On the boundary of the grid, we set $\bar{\ww}_\jj = \ww(\xx_\jj)$ since we can only interpolate to $\xx_\ii$ if all of its neighbors have data. This yields a square, symmetric positive definite system of equations for the remaining $\bar{\ww}_\jj$. The system is very well conditioned with sparse, symmetric matrix $N_\jj(\xx_\ii)$ consisting of non-negative entries and rows that sum to one.  The sparsity and symmetry of the system arises from the compact support and geometric symmetry, respectively, of the B-spline basis functions $N_\jj$. The system can be solved to a residual of machine precision in one iteration of PCG (or tens of iterations of unpreconditioned CG). In practice, we have noticed that for some flows, determining the coefficients $\bar{\ww}_\jj$ can lead to increasingly oscillatory velocity fields. This is perhaps due to the unfavorable filtering properties of even order B-splines  \cite{cheng:2001:quadratic,cheney:2012:numerical}. However, we found that a simple stabilization strategy can be obtained as
\begin{align}\label{eq:BSLQBsys}
\sum_\jj \left(\lambda N_\jj(\xx_\ii) + (1-\lambda)\delta_{\ii\jj}\right)\bar{\ww}_\jj=\ww(\xx_\ii)
\end{align}
where $\lambda\in[0,1]$ and $\delta_{\ii\jj}$ is the Kronecker delta. A value of $\lambda=0$ is very stable, but extremely dissipative. Stable yet energetic behavior is achieved by decreasing the value of $\lambda$ under grid refinement. In practice we found that $\lambda\in (.95,1 ]$ with $\lambda =c\Delta x$ for constant $c$ provided a good balance without compromising second order accuracy of the method (see Section~\ref{sec:ex_hybrid}). We note that Riish{\o}jgaard et al. \shortcite{riishojgaard:1998:use} also added diffusion to cubic spline interpolation based semi-Lagrangian to reduce noise.
\subsection{Hybrid BSLQB-PolyPIC Advection}\label{sec:poly}
In some portions of the domain, we store particles with positions $\xx_p^n$ and PolyPIC \cite{fu:2017:poly} velocity coefficients $\cc^n_p$. In the vicinity of the particles, we use PolyPIC \cite{fu:2017:poly} to update the intermediate velocity field $\bar{\ww}_\jj$. First we update particle positions as $\xx^{n+1}_p=\xx^{n}_p + \Delta t \vv_p^{n}$ (where the velocity $\vv_p^{n}$ is determined from $\cc^n_p$ following \cite{fu:2017:poly}). Then the components $\bar{w}_{\jj\alpha}$ of the coefficients $\bar{\ww}_\jj$ are determined as
\begin{align}\label{eq:polyp2g}
\bar{w}_{\jj\alpha}=\frac{\sum_p m_pN_\jj(\xx_p^{n+1})\left(\sum_{r=1}^{N_r} s_r(\xx_\jj-\xx^{n+1}_p)c^n_{pr\alpha}\right)}{\sum_p m_pN_\jj(\xx_p^{n+1})}
\end{align}
where $N_r$ is the number of polynomial modes $s_r(\xx)$, as in Fu et al. \shortcite{fu:2017:poly}. To create our hybrid approach, we update  $\bar{w}_{\jj\alpha}$ from Equation~\eqref{eq:polyp2g} whenever the denominator is greater than a threshold $\sum_p m_pN_\jj(\xx_p^{n+1})>\tau^m$, otherwise we use the BSLQB update from Equation~\eqref{eq:BSLQBsys}. We use this threshold because the grid node update in Equation\eqref{eq:polyp2g} loses accuracy when the denominator is near zero and in this case the BSLQB approximation is likely more accurate. Note that the polynomial mode coefficients for the next time step $\cc^{n+1}_p$ are determined from the grid velocities at the end of the time step (using particle positions $\xx_p^{n+1}$ and after pressure projection).

\section{Pressure Projection}
\begin{figure}[!ht]
\centering
\begin{tabular}{cc}
\subf{\includegraphics[draft=\mydraft,width=.45\columnwidth]{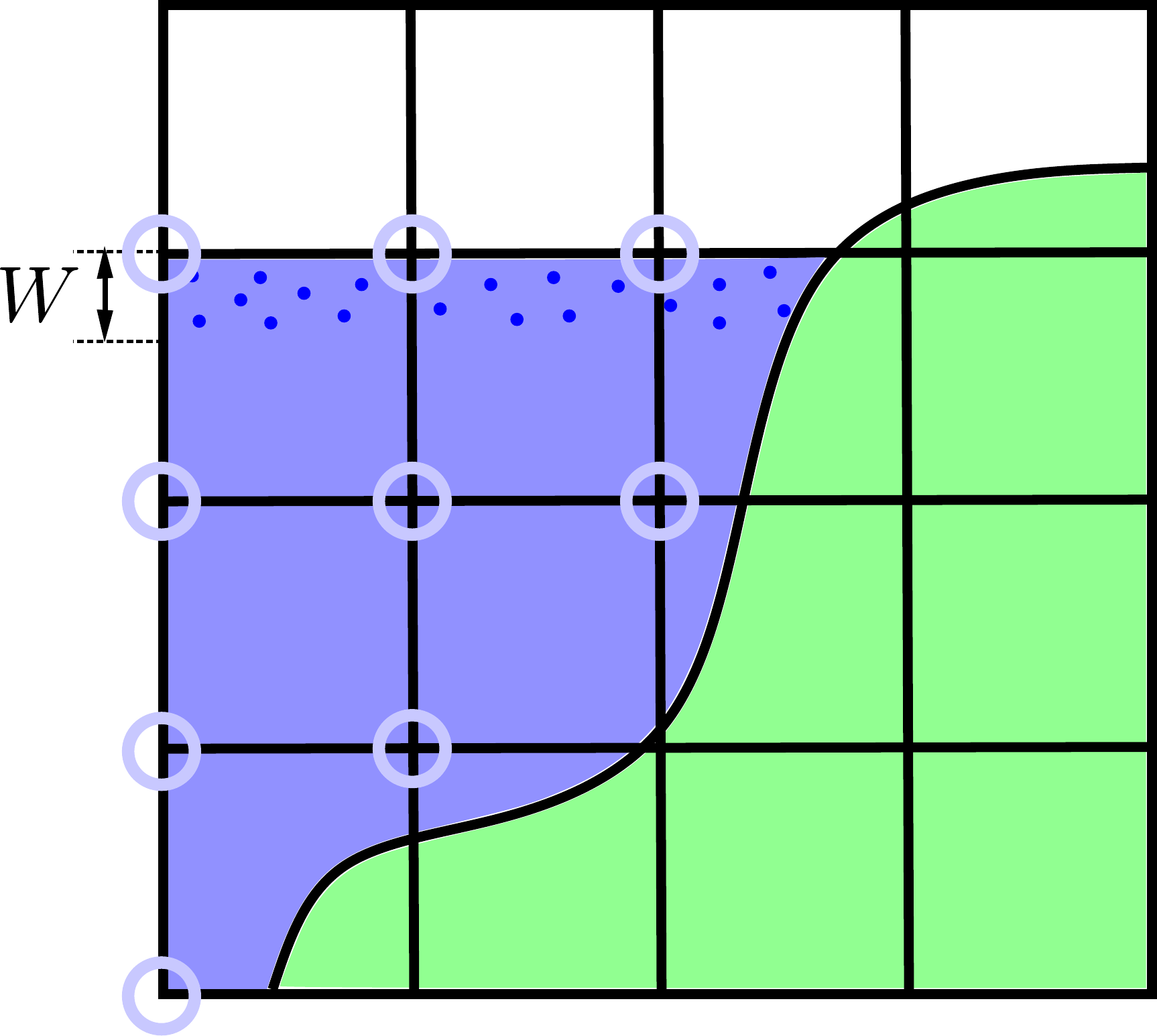}}
&
\subf{\includegraphics[draft=\mydraft,width=.4\columnwidth]{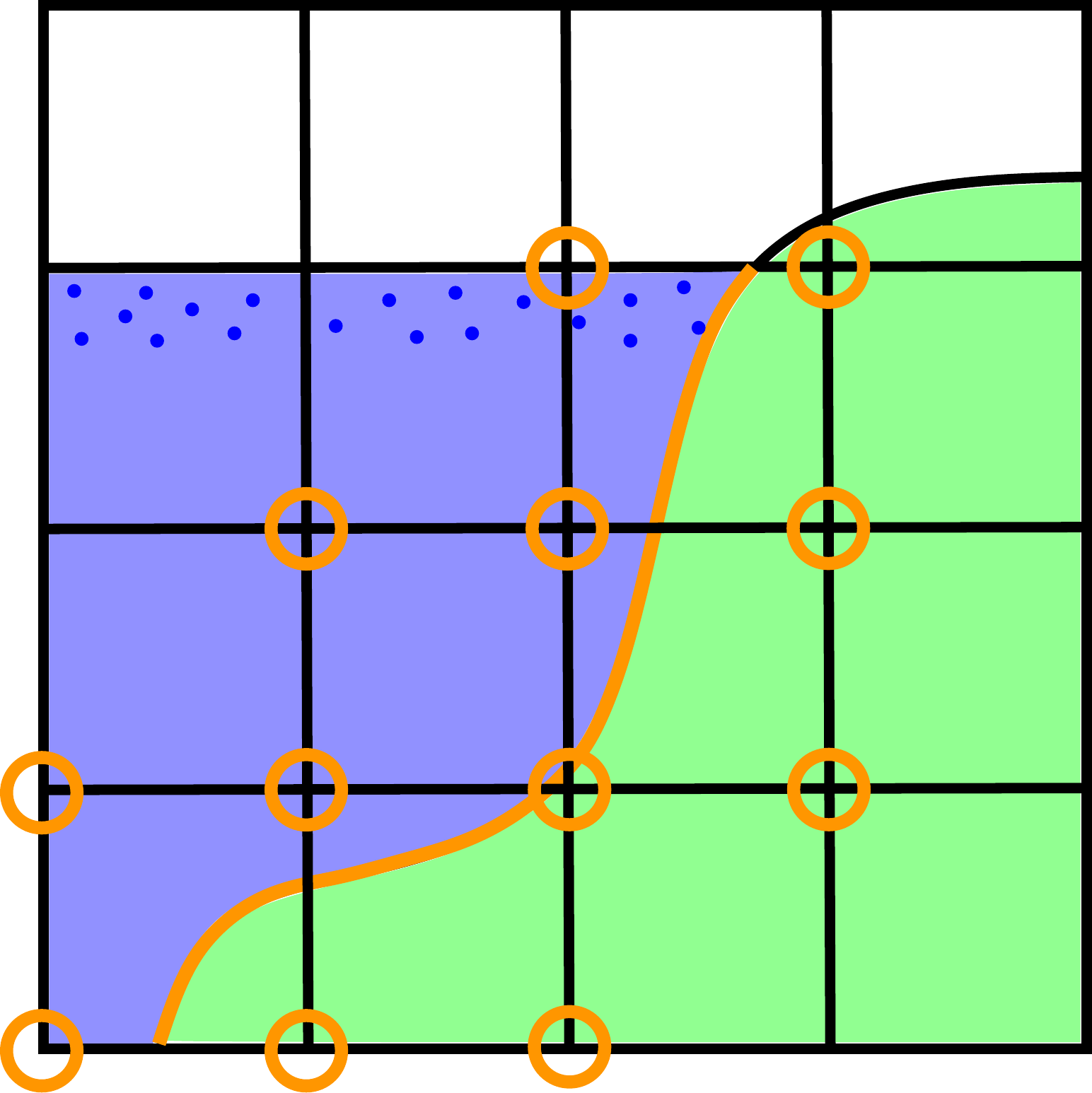}}
\\
\end{tabular}
\caption{{\textbf{Discrete free surface fluid domain}}. {\textbf{Left}}: We define the fluid domain to consist of cells that either have (1) a particle (dark blue) in it or (2) a node with non-positive level set value (light blue). {\textbf{Right}}: Boundary Lagrange multiplier external pressure $\lambda_\bb$ (orange circles) are like the interior pressures $p_\cc$ except only defined on fluid domain cells that intersect $\partial \Omega_D$.}\label{fig:fsdnb}
\end{figure}

We solve Equations~\eqref{eq:split_p}-\eqref{eq:split_div} and boundary condition Equations~\eqref{eq:bcv_cont}-\eqref{eq:bcp_cont} in a variational way. To do this, we require that the dot products of Equations~\eqref{eq:split_p}, \eqref{eq:split_div} and Equations~\eqref{eq:bcv_cont} with arbitrary test functions $\rr$, q and $\mu$ respectively integrated over the domain are always equal to zero. The free surface boundary condition in Equation~\eqref{eq:bcp_cont} is naturally satisfied by our treatment of Equation~\eqref{eq:split_p}. We summarize this as 
\begin{align}
\int_\Omega \rr\cdot\rho\left(\frac{\uu^{n+1}-\ww}{\Delta t}\right)d\xx&=\int_\Omega p^{n+1}\nabla\cdot\rr + \rho\rr\cdot \gg d\xx \label{eq:var_mom}\\
&-\int_{\partial \Omega}p^{n+1} \rr\cdot \nn ds(\xx)\nonumber\\
\int_{\Omega} q \nabla \cdot \uu^{n+1} d\xx&=0 \label{eq:var_div}\\
\int_{\partial \Omega_D} \mu \left(\uu^{n+1}\cdot\nn-a\right) ds(\xx)&=0.\label{eq:var_bcv}
\end{align}
Here we integrate by parts in the integral associated with Equation~\eqref{eq:split_p}. Furthermore, we modify the expression $\int_{\partial \Omega}p^{n+1} \rr\cdot \nn ds(\xx)$ in Equation~\eqref{eq:var_mom} in accordance with the boundary conditions. We know that the pressure is zero on $\partial \Omega_N$, however we do not know its value on $\partial \Omega_D$. We introduce the pressure on this portion of the domain as a Lagrange multiplier $\lambda^{n+1}$ associated with satisfaction of the velocity boundary condition in Equation~\eqref{eq:var_bcv}. Physically, this is the external pressure we would need to apply on $\partial \Omega_D$ to ensure that $\uu^{n+1}\cdot\nn=a$. With this convention, we have $\int_{\partial \Omega}p^{n+1} \rr\cdot \nn ds(\xx)=\int_{\partial \Omega_D}\lambda^{n+1} \rr\cdot \nn ds(\xx)$. We note that unlike Equation~\eqref{eq:var_bcv} (and its strong form counterpart $\eqref{eq:bcv_cont}$) that requires introduction of a Lagrange multiplier, Equation~\eqref{eq:bcp_cont} is naturally enforced through the weak form simply by setting $p^{n+1}=0$ in the integral over $\partial \Omega_N$ in Equation~\eqref{eq:var_mom}.\\
\\
To discretize in space, we introduce interpolation for the test functions $\rr$, $q$ and $\mu$. We use the same spaces as in Equation~\eqref{eq:interp} for velocity and pressure for $\rr=\sum_\ii \bar{\rr}_\ii N_\ii$ and $q=\sum_\dd q_\dd \chi_\dd$. For the test functions $\mu$, we choose the same space as $q,p$, but with functions restricted to $\partial \Omega_D$, $\mu=\sum_\bb \mu_\bb \chi_\bb$ for $\bb$ with grid cell $\Omega_\bb \cap \partial \Omega_D \neq \emptyset$ (see Figure~\ref{fig:fsdnb}). We choose the same space for $\lambda^{n+1}=\sum_\bb\lambda^{n+1}_\bb \chi_\bb$ to close the system. With these choices for the test functions, the variational problem is projected to a finite dimensional problem defined by the interpolation degrees of freedom. This is expressed as a linear system for velocities $\bar{\uu}_\jj^{n+1}$, internal pressures $p^{n+1}_\cc$, and external pressures $\lambda_\bb^{n+1}$ that is equivalent to 
\begin{align}
\left(\begin{array}{ccc}
\MM&-\DD^T&\BB^T\\
-\DD&&\\
\BB&&
\end{array}\right)
\left(
\begin{array}{c}
\UU^{n+1}\\
\PP^{n+1}\\
\boldsymbol\Lambda^{n+1}
\end{array}
\right)=
\left(
\begin{array}{c}
\MM\WW  + \hat{\gg}\\
\mb{0}\\
\AA
\end{array}
\right).
\end{align}
Here $\UU^{n+1}$, $\PP^{n+1}$ and $\boldsymbol\Lambda^{n+1}$ are the vectors of all unknown $\bar{\uu}_\jj^{n+1}$, $p_\cc^{n+1}$ and $\lambda_\bb^{n+1}$ respectively. Furthermore $\MM$ is the mass matrix, $\BB$ defines the velocity boundary conditions and $\DD$ defines the discrete divergence condition. Lastly, $\WW $ is the vector of all $\bar{\ww}_\ii$ that define the intermediate velocity, $\hat{\gg}$ is from gravity and $\AA$ is the variational boundary condition. Using the convention that Greek indices $\alpha,\beta$ range from $1-3$, these matrices and vectors have entries 
\begin{align}
M_{\alpha\ii\beta\jj}=\delta_{\alpha\beta}\int_\Omega \frac{\rho}{\Delta t} N_\ii N_\jj d\xx, \ D_{\dd \beta \jj}&=\int_\Omega \chi_\dd\frac{\partial N_\jj}{\partial x_\beta}d\xx, \ \hat{g}_{\alpha \ii}=\int_\Omega \rho g_\alpha N_\ii d\xx \label{eq:vol_int} \\
B_{\bb \beta\jj}=\int_{\Omega_D} \chi_\bb N_\jj n_\beta ds(\xx), \ A_\bb &= \int_\Omega a\chi_\bb ds(\xx).\label{eq:b_int}
\end{align}
If we define $\GG=[-\DD^T,\BB^T]$, we can convert this system into a symmetric positive definite one for $\PP^{n+1}$ and $\boldsymbol\Lambda^{n+1}$ followed by a velocity correction for $\UU^{n+1}$
\begin{align}
\left(
\begin{array}{c}
\label{eq:spd_system}
\PP^{n+1}\\
\boldsymbol\Lambda^{n+1}
\end{array}
\right)&=\left(\GG^{T}\MM^{-1}\GG\right)^{-1}
\left(\GG^T\left(\WW+\MM^{-1}\hat{\gg}\right)-\left(\begin{array}{c}\mb{0}\\\AA\end{array}\right)\right)\\
\UU^{n+1}&=-\MM^{-1}\GG\left(
\begin{array}{c}
\PP^{n+1}\\
\boldsymbol\Lambda^{n+1}
\end{array}
\right)+\WW+\MM^{-1}\hat{\gg}.
\end{align}
Unfortunately, this system will be dense in the current formulation since the full mass matrix $M_{\alpha\ii\beta\jj}$ is non-diagonal with dense inverse \cite{botella:2002:collocation}. However, a simple lumped mass approximation
\begin{align}\label{eq:mass_lump}
M^l_{\alpha\ii\beta\jj}=\left\{
\begin{array}{lcc}
\delta_{\alpha\beta}\int_\Omega \frac{\rho}{\Delta t} N_\ii d\xx,&\ii=\jj\\
0,&\textrm{otherwise}
\end{array}
\right.
\end{align}
gives rise to a sparse matrix in Equation~\eqref{eq:spd_system}.

\begin{figure}[!ht]
  \centering
  \begin{subfigure}{.49\columnwidth}
     \includegraphics[draft=\mydraft,clip,width=1.0\columnwidth]{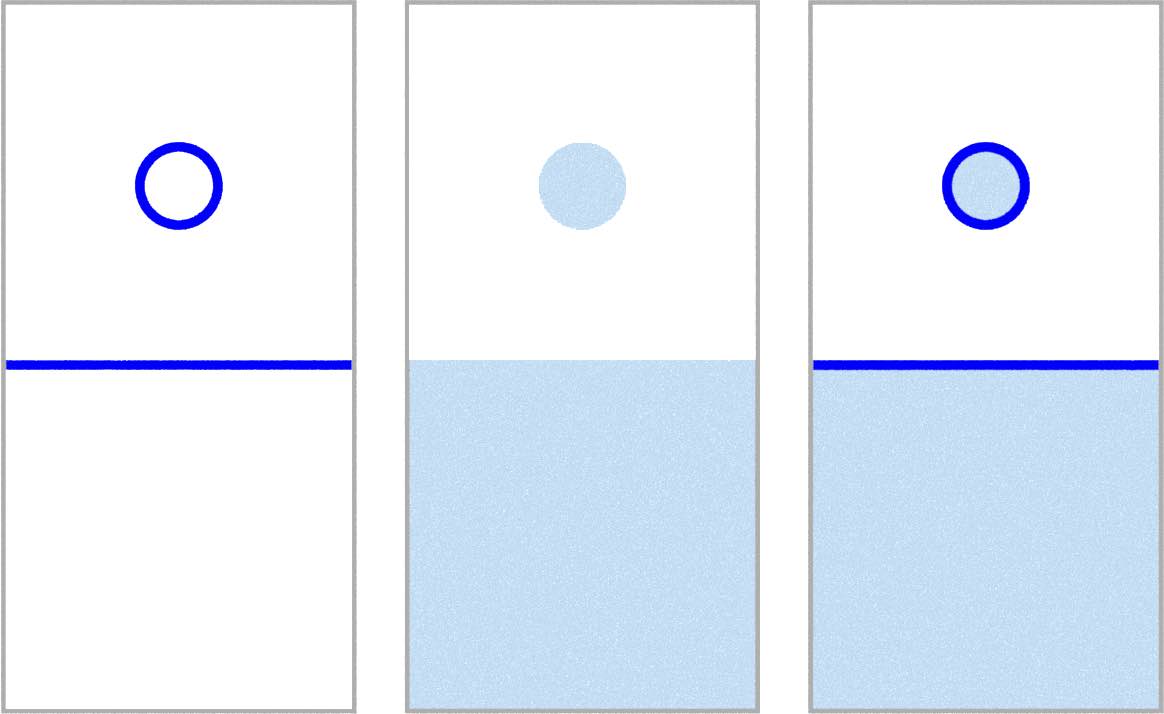}
  \end{subfigure} 
  \begin{subfigure}{.49\columnwidth}
     \includegraphics[draft=\mydraft,clip,width=1.0\columnwidth]{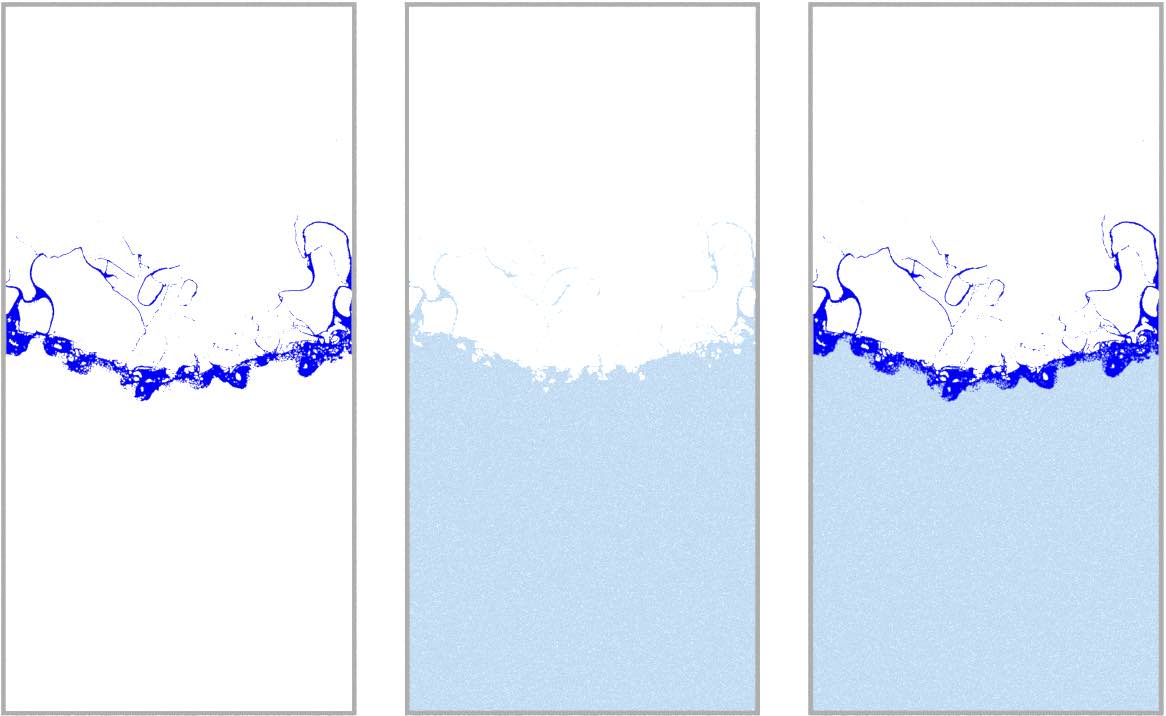}
  \end{subfigure}\\
  \begin{subfigure}{0.33\columnwidth}
     \includegraphics[draft=\mydraft,trim={920px 430px 830px 470px},clip,width=1.0\columnwidth]{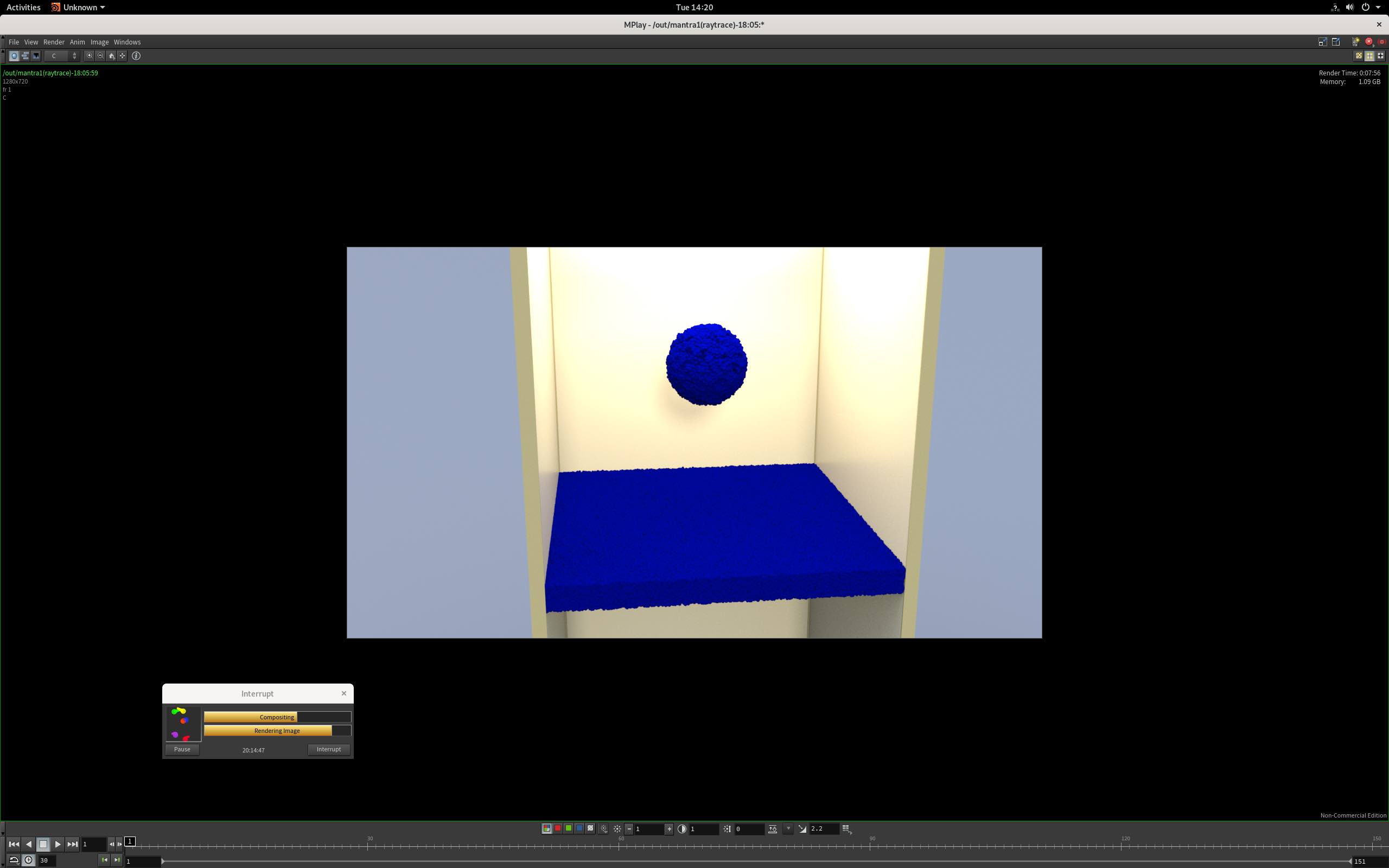}
  \end{subfigure}
  \begin{subfigure}{0.33\columnwidth}
     \includegraphics[draft=\mydraft,trim={920px 430px 830px 470px},clip,width=1.0\columnwidth]{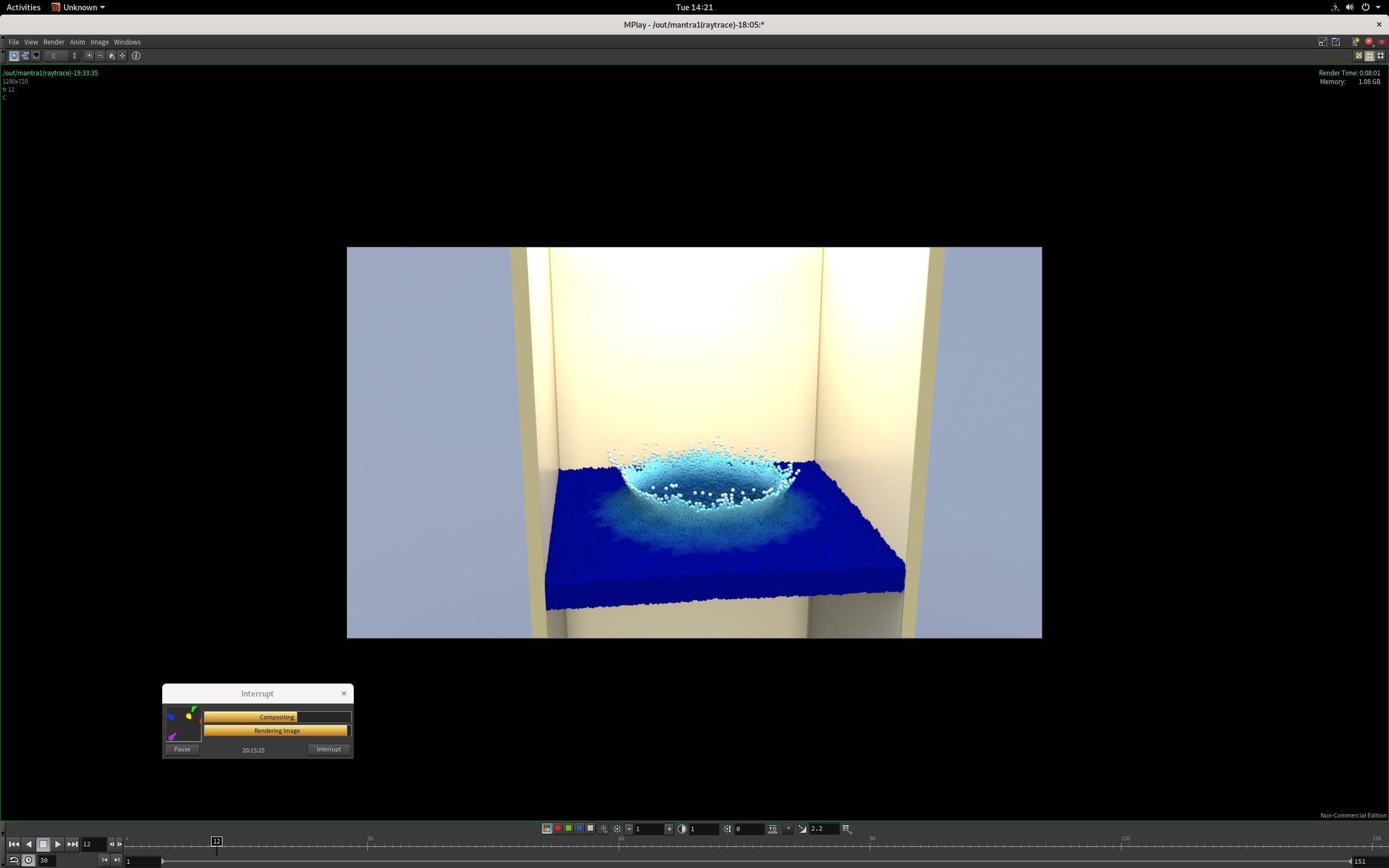}
  \end{subfigure} 
  \begin{subfigure}{0.33\columnwidth}
     \includegraphics[draft=\mydraft,trim={920px 430px 830px 470px},clip,width=1.0\columnwidth]{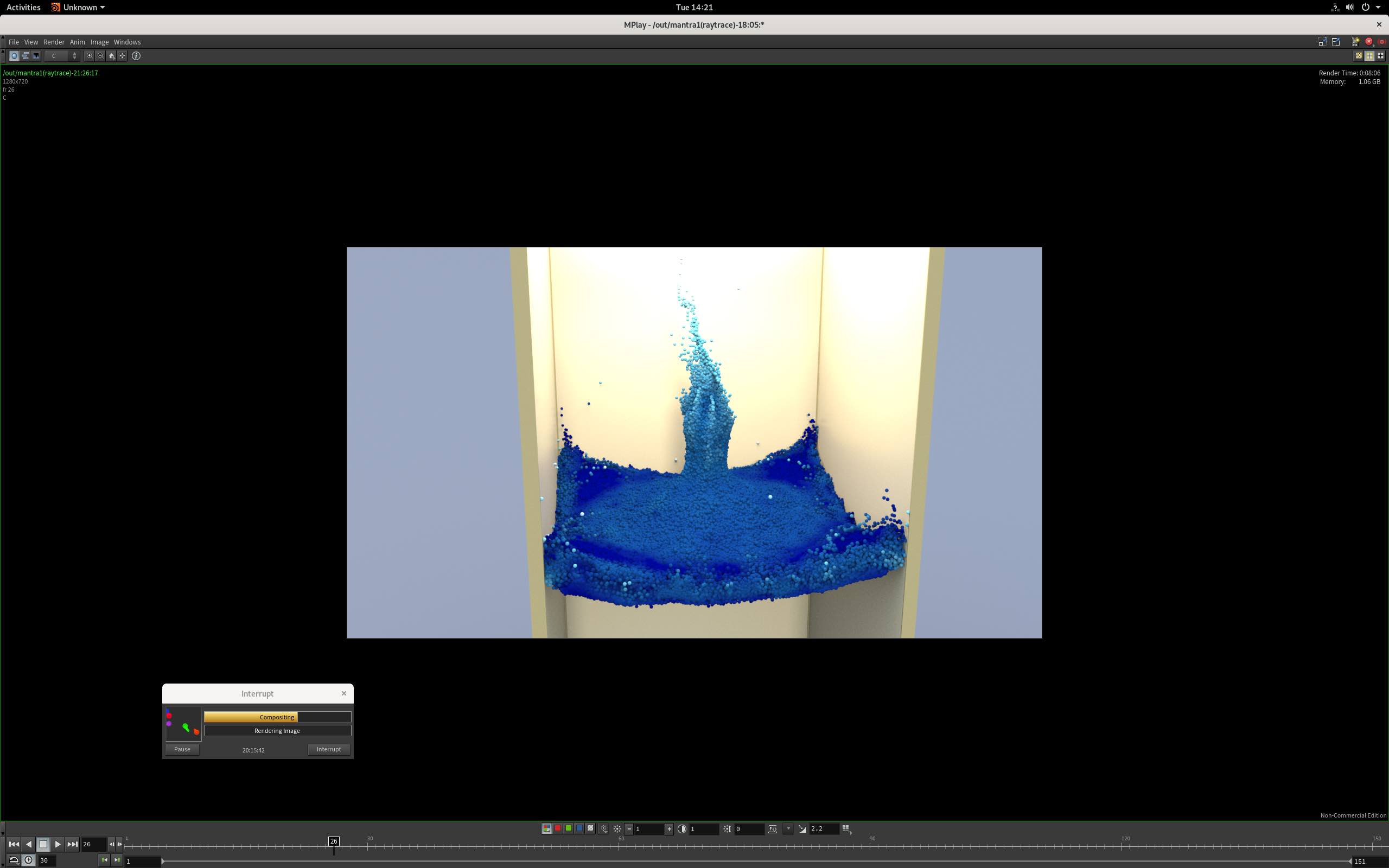}
  \end{subfigure}
\caption{{\textbf{Narrow band free surface}}. A circle/sphere falls in a tank of water under gravity. Using only a narrow band of particles saves computational cost and enables increased resolution of the free surface. {\textbf{Top}}: In 2D we illustrate the hybrid particle(dark blue)/level set (light blue) representation. {\textbf{Bottom}}: Particles are colored based on velocity magnitude.}\label{fig:narrow_band_2d}
\end{figure}


\subsection{Cut cells}
\label{sec:cutcells}
\begin{figure}[!ht]
\includegraphics[draft=\mydraft,width=\columnwidth]{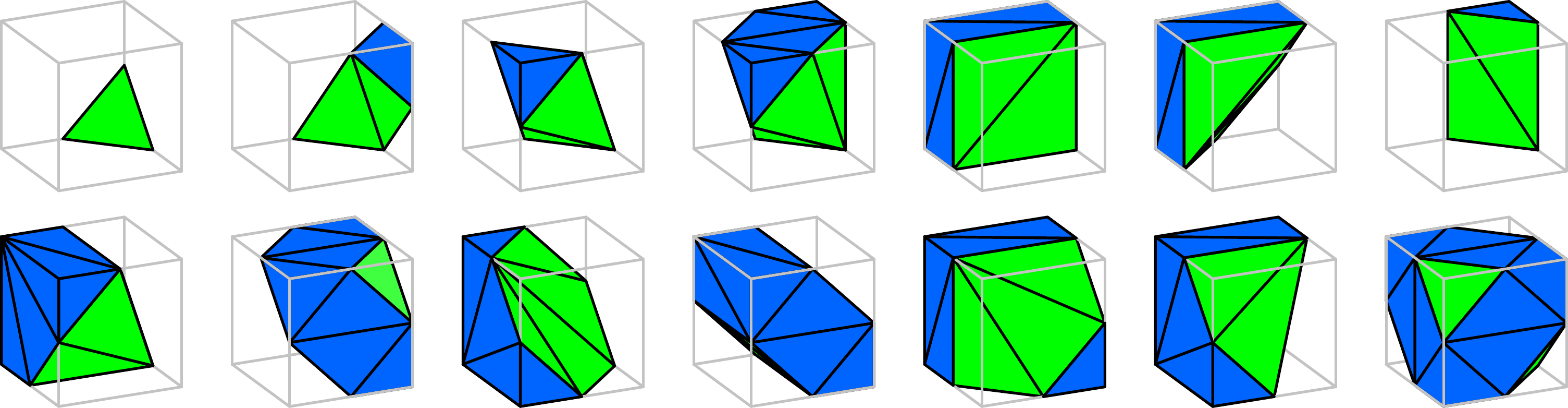}
\caption{{\textbf{Cut cells}}. We show the 14 essential cases used in determining the cut cell fluid domain geometry. Blue faces indicate the intersection of the grid cell with the fluid domain. Green faces indicate the velocity boundary condition faces on $\partial \Omega_D$.}\label{fig:cutcell}
\end{figure}

As in XFEM and VNA approaches \cite{belytschko:2009:review,koschier:2017:xfem,schroeder:2014:vna}, we resolve sub-grid-cell geometry by simply performing the integrations in Equations~\eqref{eq:vol_int}-\eqref{eq:b_int} over the geometry of the fluid domain. We use a level set to define solid boundaries (green in Figure~\ref{fig:fsdnb}) on which velocity boundary conditions are defined. We triangulate the zero isocontour using marching cubes \cite{chernyaev:1995:marching} (see Figure~\ref{fig:cutcell}). The integrals in Equations~\eqref{eq:vol_int}-\eqref{eq:b_int} all involve polynomials over volumetric polyhedra (Equations~\eqref{eq:vol_int}, blue in Figure~\ref{fig:cutcell}) or surface polygons (Equations~\eqref{eq:b_int}, green in Figure~\ref{fig:cutcell}) and we use Gauss quadrature of order adapted to compute the integrals with no error (see \cite{gagniere:2020:tech_doc}). For free surface flows, we use particles (and additionally a level set function in the case of narrow banding, see Section~\eqref{sec:nb}) to denote grid cells with fluid in them. Cells near the solid boundary are clipped by the marching cubes geometry. The fluid domain $\Omega$ is defined as the union of all clipped and full fluid cells (see Figure~\ref{fig:fsdnb}). \\ 
\\
Notably, taking a cut cell approach with our variational formulation allows us to prove that our method can resolve a standing pool of water exactly without producing numerical currents. We know that with gravitational force $\rho\gg$ (e.g. with $\gg$ pointing in the $y$ direction with magnitude $g$), steady state is maintained if the pressure increases with depth as $p=\rho g\left(y_0-y\right)$ where $y_0$ is the height of the water surface at rest, since $-\nabla p + \rho\gg = \mb{0}$. Since we use multilinear interpolating functions for $p$, the exact solution is representable in our discrete space and with a short proof we show (see \cite{gagniere:2020:tech_doc}) that this means our method will choose it to maintain a standing pool of water, independent of fluid domain boundary geometry.

\section{Narrow band free surface}\label{sec:nb}
For free surface flows, we develop a narrow band approach as in \shortcite{chentanez:2015:coupling,ferstl:2016:narrow,sato:2018:nb}. We represent the fluid domain with a level set and seed particles in a band of width $W$ from the zero isocontour (see Figure~\ref{fig:fsdnb}). Particles are advected and used to augment BSLQB advection as detailed in Section~\ref{sec:poly}. We also advect the level set by interpolating its value at the previous step from the upwind location $\xx_\ii-\Delta t \ww(\xx_\ii)$ determined in Equation~\eqref{eq:ad_disc}. We then use the updated particle locations to compute a narrow band level set from the particles based on the method of Boyd and Bridson \cite{boyd:2012:multiflip}. We update the level set to be the union of that defined by the narrow band and that from advection. This is done by taking the minimum of the two level set values and then redistancing with the method of Zhao \shortcite{zhao:2005:fast}.
\section{Examples}
\begin{figure}[h]
\includegraphics[draft=\mydraft,width=\columnwidth]{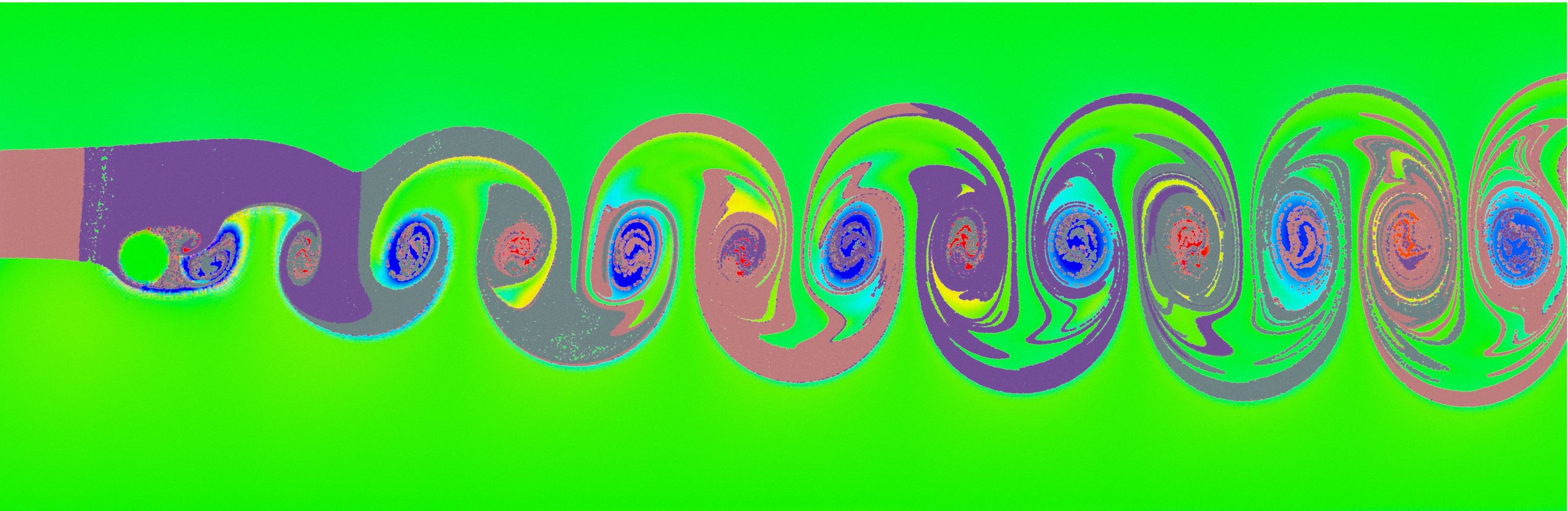}
\caption{{\textbf{Von Karman vortex shedding}}. We demonstrate the accuracy of our Hybrid BSLQB/PolyPIC with vortex shedding past a notch in 2D. Note the smooth transition between regions with particles (PolyPIC) and those without (BSLQB).}\label{fig:vonK}
\end{figure}


\subsection{Hybrid BSLQB/PolyPIC}\label{sec:ex_hybrid}
We demonstrate our hybrid BSLQB/PolyPIC advection with water simulation. We prevent excessive run times by utilizing a narrow band of particles near the free surface and a level set (with BSLQB advection) in deeper levels. Figure~\ref{fig:narrow_band_2d} Top shows a disc of water splashing in a rectangular tank with dimension $1\times2$ and grid cell size $\Delta x = 1/255$. The time step is restricted to be in the range $\Delta t \in \left[0.005, 0.01\right]$.  20 particles are initialized in every cell that is initially in a narrow band of $7 \Delta x$ below the zero isocontour of the level set. Figure~\ref{fig:narrow_band_2d} Bottom shows an analogous 3D example where a sphere of water splashes in a tank.  A cell size of $\Delta x = \frac{1}{63}$ is used in a domain with dimensions $1\times 2 \times 1$.  We take a fixed time step of $\Delta t = 0.01$ and demonstrate that narrow banding does not prevent larger-than-CFL time steps.  1,008,187 particles are used to resolve the free surface in a narrow band of width $5\Delta x$.  As in 2D, the particles capture highly-dynamic behavior of the free surface while the level set is sufficient to represent the bulk fluid in the bottom half of the domain.\\
\\
We also demonstrate our hybrid advection with a vortex shedding example (see Figure~\ref{fig:vonK}). The flow domain $\Omega$ is a $3\times 1$ rectangle with circle of radius $0.05$. We seed a band of particles of width $.2$ above the midline $y=.5$ for PolyPIC advection. Advection in the rest of the domain is done with BSLQB. The vorticity plot illustrates a seamless transition between the two advection schemes. The simulation was run with a grid resolution of $\Delta x=\frac{1}{255}$, CFL number of 4 (i.e. $\Delta t = \frac{4\Delta x}{v_\textrm{max}}$), and inlet speed of $1.5$.
\begin{figure}[!b]
  \centering
  \begin{subfigure}{.49\columnwidth}
     \includegraphics[draft=\mydraft,clip,width=\columnwidth]{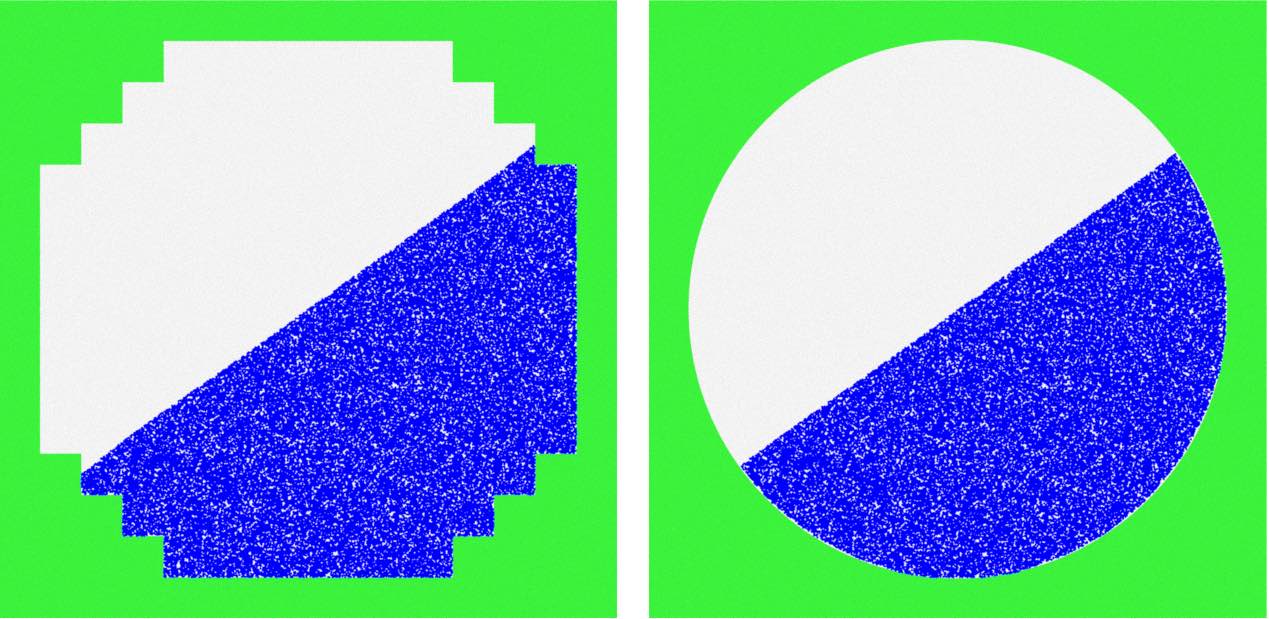}
  \end{subfigure}
  \begin{subfigure}{.49\columnwidth}
     \includegraphics[draft=\mydraft,clip,width=\columnwidth]{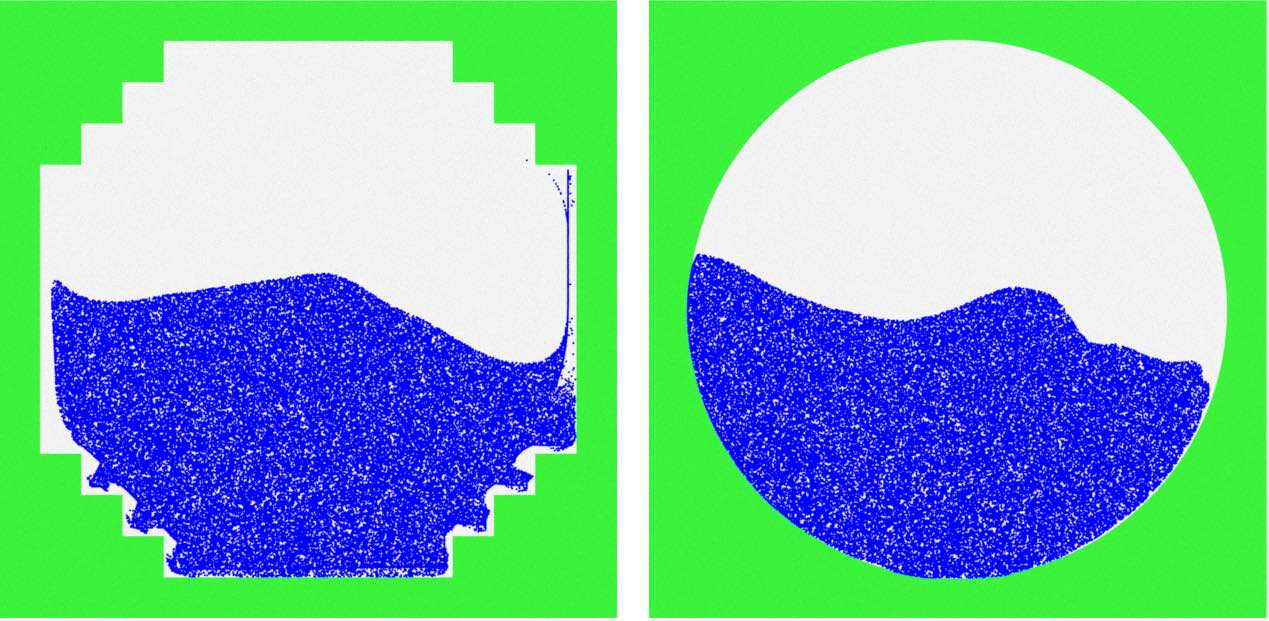}
  \end{subfigure}
\caption{{\textbf{Cut cell vs.\ voxelized domain}}. Using a cut cell domain (right) instead of a voxelized domain (left) yields marked improvements in simulation quality.}\label{fig:voxelized_vs_cut_cell}
\end{figure}

\subsection{BSLQB comparison with explicit semi-Lagrangian}
\begin{figure}[ht]
  \centering
  \begin{subfigure}{1.0\columnwidth}
     \includegraphics[draft=\mydraft,trim={0 60px 0 60px},clip,width=\columnwidth]{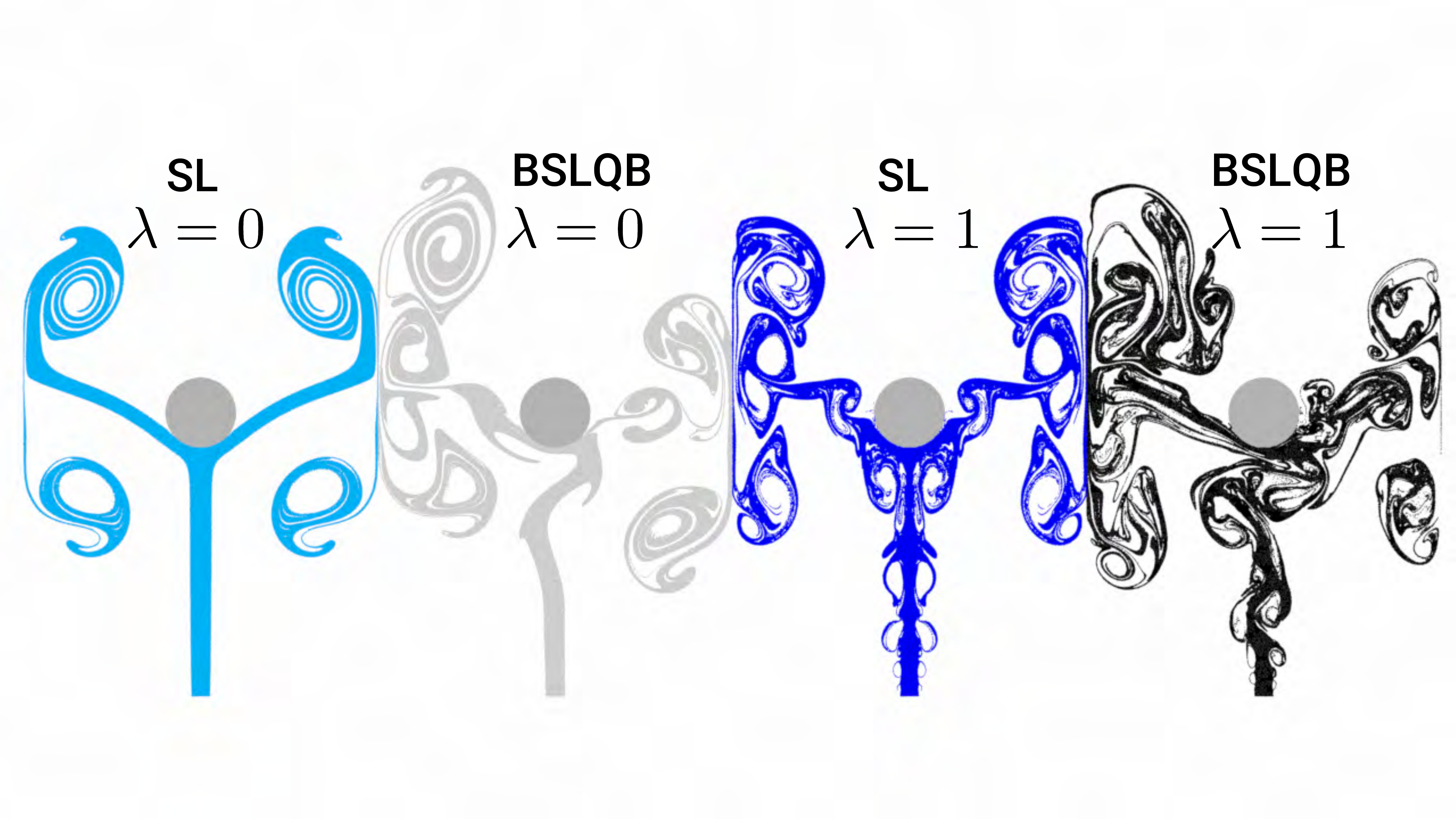}
  \end{subfigure}
  \\
  \begin{subfigure}{1.0\columnwidth}
     \includegraphics[draft=\mydraft,trim={0 40px 0 300px},clip,width=\columnwidth]{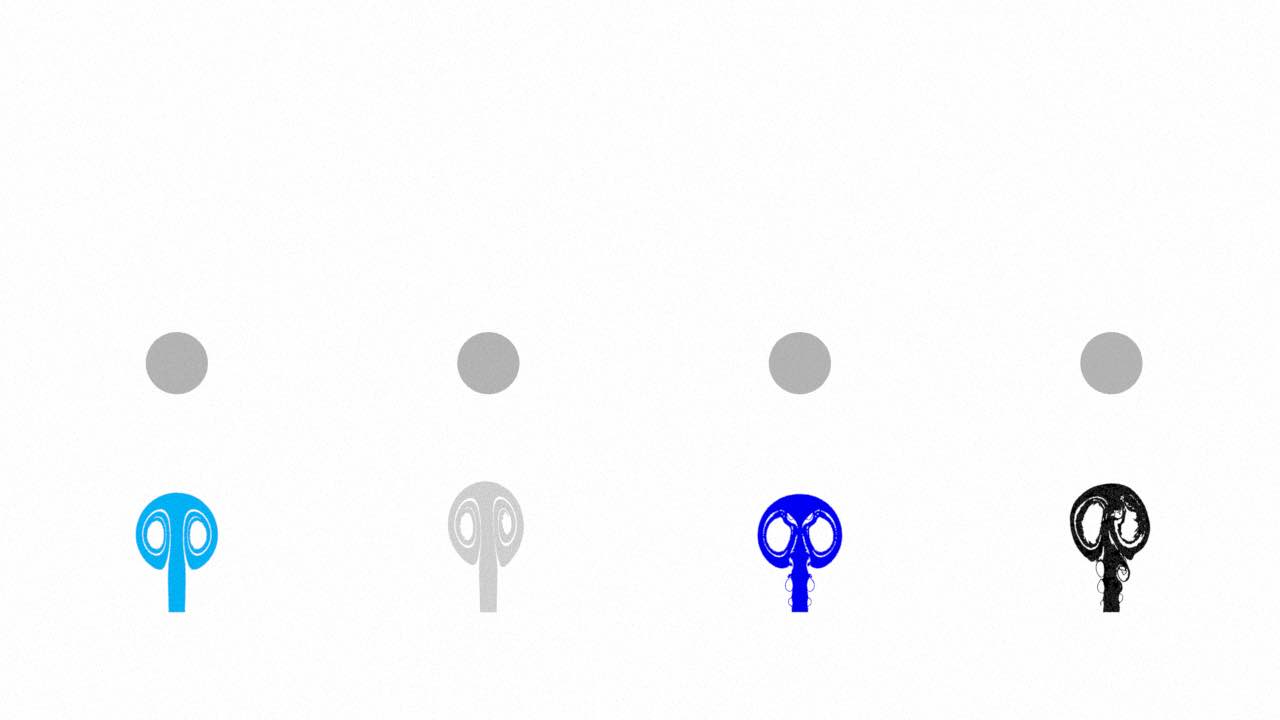}
  \end{subfigure}
\caption{{\textbf{Interpolation correction}}. BSLQB exhibits more fine-scale flow detail and vorticity than semi-Lagrangian for extremal values of interpolation parameter $\lambda$ (Equation~\eqref{eq:BSLQBsys}). From left to right: semi-Lagrangian with $\lambda = 0$, BSLQB with $\lambda = 0$, semi-Lagrangian with $\lambda = 1$, BSLQB with $\lambda = 1$.}\label{fig:interp_correction}
\end{figure}

\begin{figure}[ht]
\includegraphics[draft=\mydraft,width=\columnwidth]{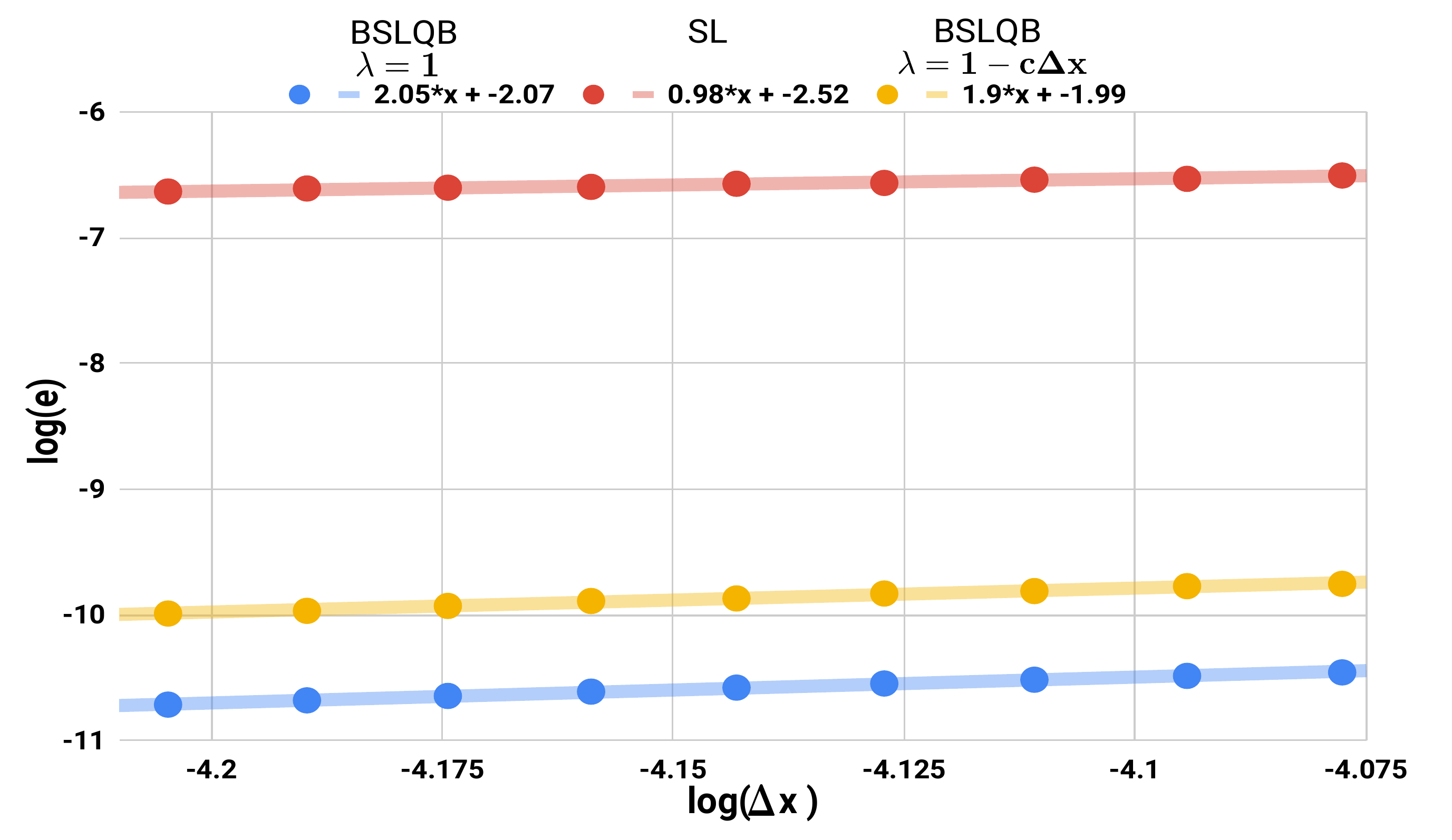}
\caption{{\textbf{Convergence}}. We compare explicit semi-Lagrangian (SL, red), with BSLQB (blue) and interpolation coefficient $\lambda=1$ (Equation~\eqref{eq:BSLQBsys}) and BSLQB with interpolation coefficient $\lambda=1-c\Delta x$ (orange). We plot $\log(\Delta x)$ versus $\log(e)$ (where $e$ is the infinity norm of the error) for a variety of grid resolutions $\Delta x$ and compute the best fit lines. The slope of the line provides empirical evidence for the convergence rate of the method.}\label{fig:conv}
\end{figure}

We demonstrate improved resolution of flow detail with BSLQB compared to explicit semi-Lagrangian in a 2D example of smoke flowing past a circle (see Figure~\ref{fig:interp_correction}) and with a 2D spinning circle example (see Figure~\ref{fig:inner_circle}). Note that particles are only used for flow visualization and not for PolyPIC advection in these examples. BSLQB exhibits more energetic, turbulent flows than semi-Lagrangian advection. Notably, the BSLQB result breaks symmetry sooner. In Figure~\ref{fig:interp_correction} we also examine the effect of extremal values of the $\lambda$ parameter described in Equation~\eqref{eq:BSLQBsys}.  A zero value of $\lambda$ is quite dissipative compared to a full value of $\lambda = 1$ for both semi-Lagrangian and BSLQB.  As mentioned in Section~\ref{sec:bslqb}, we generally found that keeping $\lambda$ close to 1 provided the least dissipative behavior, while setting the value slightly less than 1 helped restore stability when necessary (one can also dynamically adjust this value over the course of a simulation). In Figure~\ref{fig:inner_circle}, we initially set the angular velocity to 4 radians per second in a circle of radius $.2$ (with $\Omega=[0,1]\times[0,1]$). The simulation is run with $\Delta x=\frac{1}{511}$ and a $\Delta t = .02$ (CFL number of 3).\\
\\
We examine the convergence behavior of BSLQB for the 2D Burgers' equation $\frac{D\uu}{Dt}=\mb{0}$ with initial data $\uu(\xx)=\xx\cdot\left(\AA\xx\right)$ for $\AA=\RR\boldsymbol\Lambda\RR^T$ for diagonal $\boldsymbol\Lambda$ with entries $1$ and $.25$ and rotation (of $.1$ radians) $\RR$ (see Figure~\ref{fig:conv}). We examine the convergence behavior under refinement in space and time with $\Delta t=\Delta x$. We compute the best fit line to the plot of the logarithm of the $L^\infty$ norm of the error versus the logarithm of $\Delta x$ for a number of grid resolutions. We observe slopes of approximately 2 for BSLQB with interpolation parameter $\lambda=1$ and $\lambda=1-c\Delta x$ (with $c = 2.95$), indicating second order accuracy in space and time under refinement. We observe slopes of approximately 1 for explicit semi-Lagrangian, indicating first order.

\begin{figure}[!ht]
  \centering
  \begin{subfigure}{\columnwidth}
     \includegraphics[draft=\mydraft,trim={0 200px 0 200px},clip,width=\columnwidth]{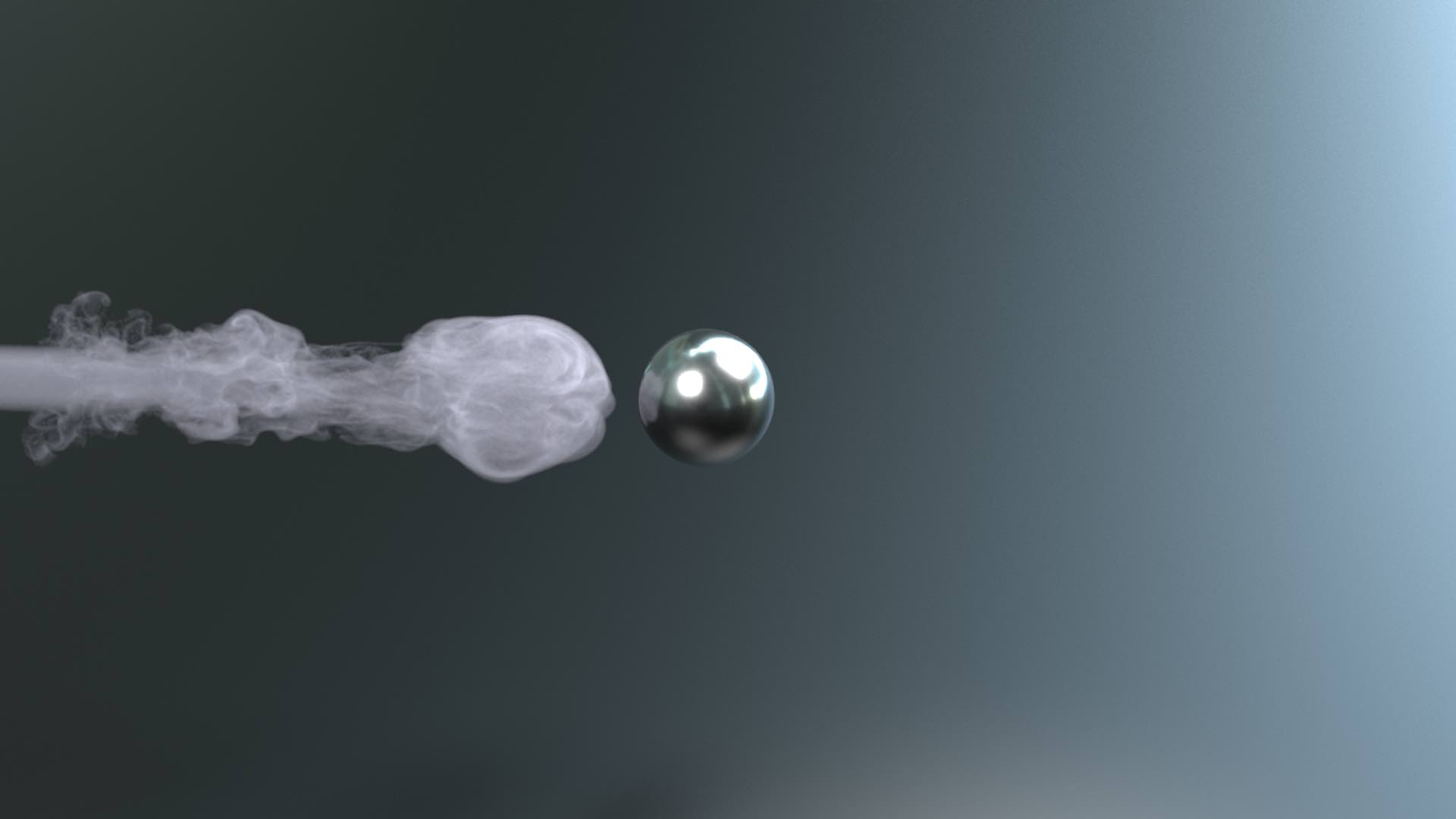}
  \end{subfigure}\\
  \begin{subfigure}{\columnwidth}
     \includegraphics[draft=\mydraft,trim={0 200px 0 200px},clip,width=\columnwidth]{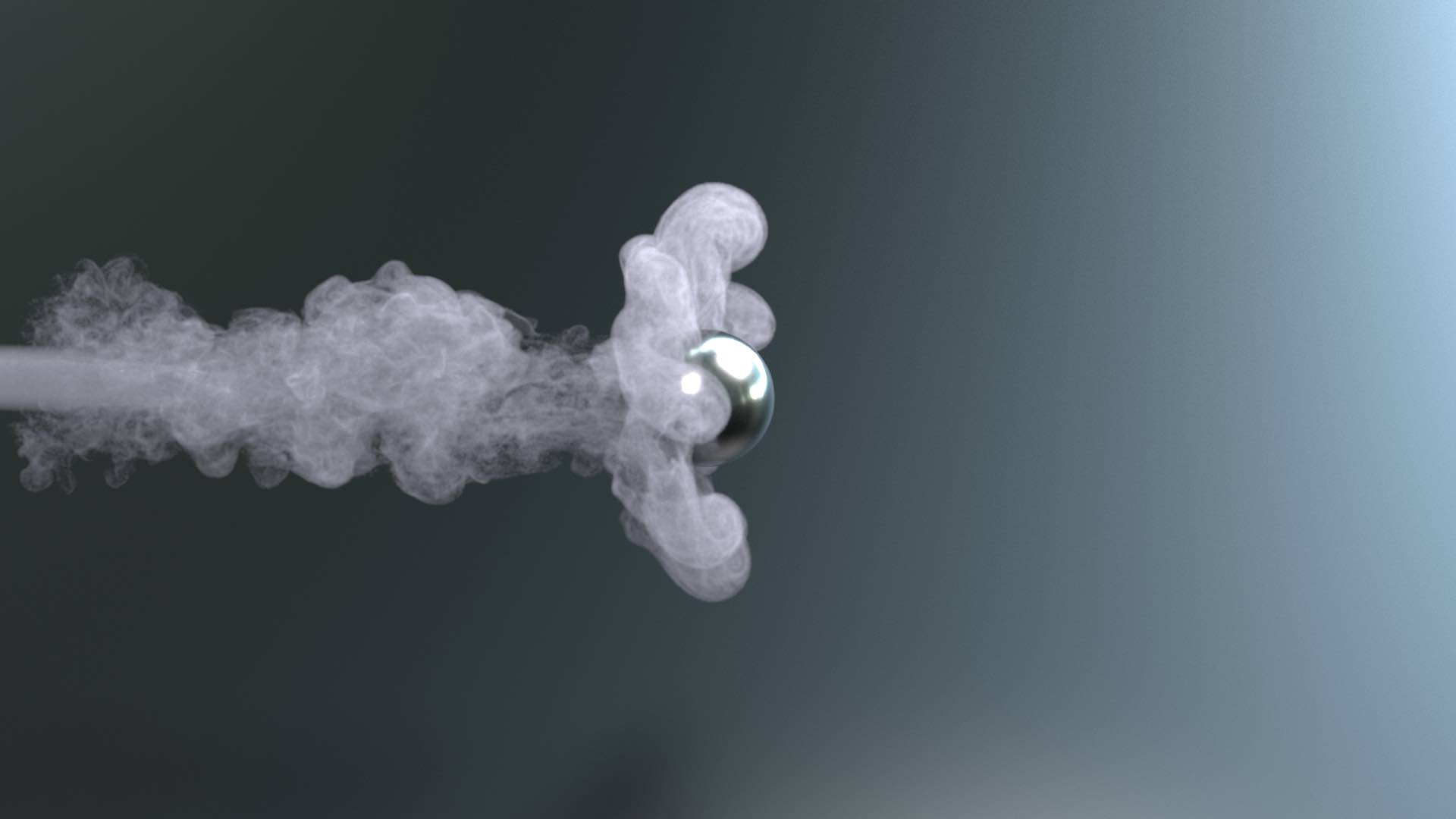}
  \end{subfigure}\\
  \begin{subfigure}{\columnwidth}
     \includegraphics[draft=\mydraft,trim={0 200px 0 200px},clip,width=\columnwidth]{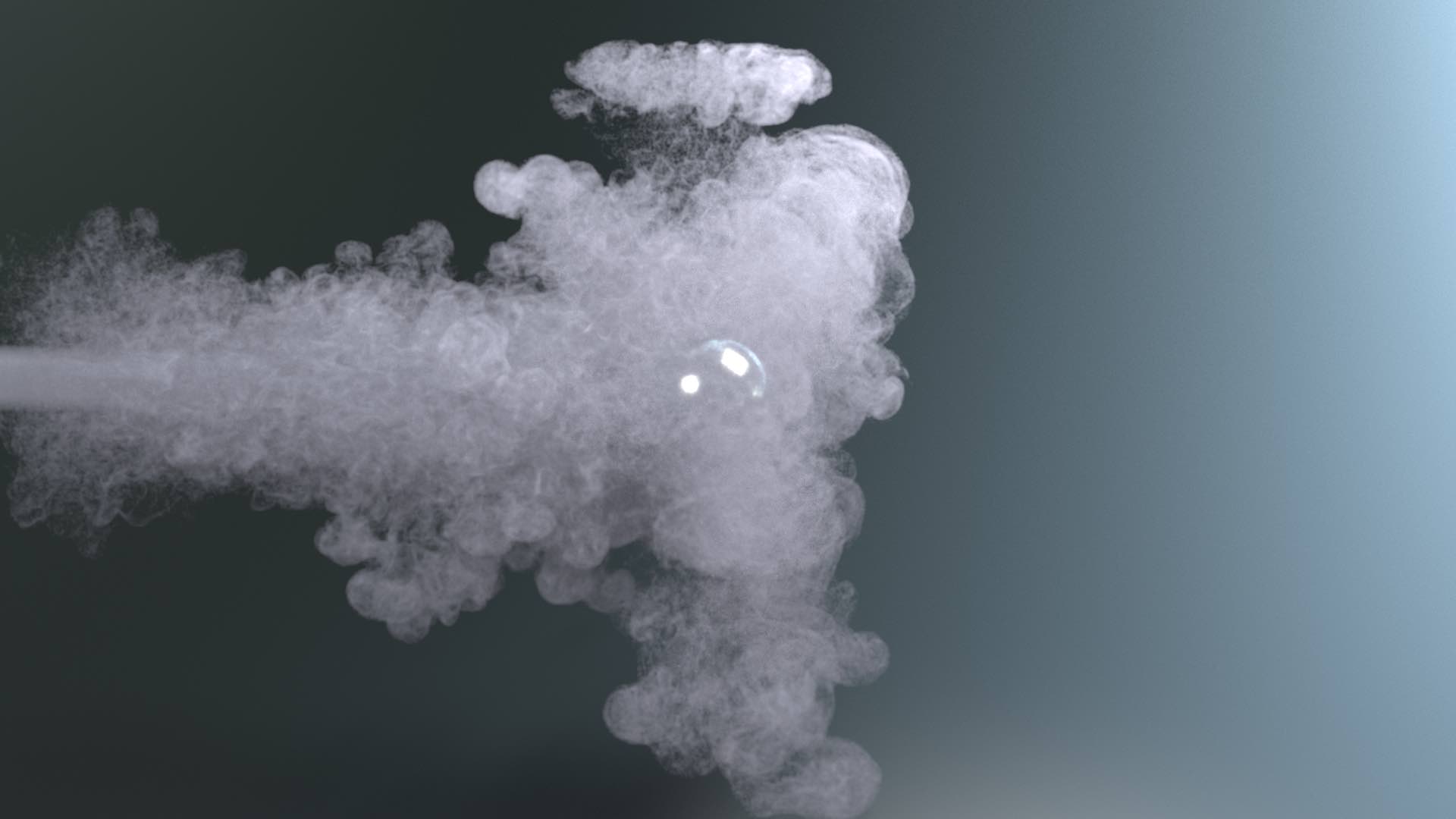}
  \end{subfigure}
\caption{{\textbf{Smoke jet}}. A plume of smoke is simulated with BSLQB. Zero normal velocity boundary conditions are enforced on the irregular boundary of the sphere inducing intricate flow patterns as the smoke approaches it.}\label{fig:smokejet}
\end{figure}

\subsection{Cut cell examples}
We demonstrate the ability of our cut cell method to produce detailed flows in complicated irregular domains for smoke and free surface water examples. Figure~\ref{fig:rainbowsmoke} demonstrates the subtle and visually interesting behavior that arises as two plumes of multicolored smoke flow to the center of a cubic domain colliding with a spherical boundary. We use $\Delta x = 1/63$ and $\Delta t = .02$. We demonstrate a more complex domain in Figure~\ref{fig:bunnysmoke}. Puffs of colored smoke with converging initial velocities are placed in a bunny shaped clear domain. We use grid size $\Delta x = 1/127$ and a fixed time step of $\Delta t = 0.01$ (CFL number $>1$). In Figure~\ref{fig:snowglobe}, we demonstrate water splashing, while accurately conforming to the walls of an irregular domain defined as the interior of a large sphere and exterior of a small inner sphere. The spatial resolution of the domain is $\Delta x = 1/127$, and 30 particles per cell are seeded in the initial fluid shape. A minimum time step of $\Delta t=0.001$ is enforced, which is often larger than the CFL condition. We also consider dam break simulations in rectangular domains with column obstacles (Figure~\ref{fig:dambreak_final}) and a bunny obstacle (Figure~\ref{fig:dambreak_final}). Both examples use a grid cell size of $\Delta x = 1/127$, 8 particles per cell and a fixed time step of $\Delta t = 0.003$. Lastly, we demonstrate the benefits of our cut cell formulation over a more simplified, voxelized approach in Figure~\ref{fig:voxelized_vs_cut_cell}. Notice the water naturally sliding in the cut cell domain compared with the jagged flow in the voxelized domain.



\subsection{Performance considerations}

\begin{table}[]
\begin{tabular}{@{}llll@{}}
\toprule
Example         & Seconds & \# Particles & $\Delta x^{-1}$   \\ \midrule
Smoke Jet (Fig.~\ref{fig:smokejet})      & 1,212             & 12,502,349 & 127   \\
Multiple Jets (Fig.~\ref{fig:rainbowsmoke})  & 53             & 25,004,699             & 63   \\
Bunny Smoke (Fig.~\ref{fig:bunnysmoke})    & 160             & 24,000,000           & 127    \\
Smoke Spheres$\text{*}$ (Fig.~\ref{fig:bigsmoke})  & 428             & 64,000,000             & 255    \\
Narrow Band (Fig.~\ref{fig:narrow_band_2d})    & 396             & 1,008,187             & 63   \\
Water Globe (Fig.~\ref{fig:snowglobe})    & 242             & 524,415             & 127  \\
Dam Break (Fig.~\ref{fig:dambreak_final})       & 870            & 3,251,409            & 127  \\
Bunny Dam Break (Fig.~\ref{fig:bunny_dambreak_final}) & 1,171             & 4,797,535           & 127 \\ \bottomrule
\end{tabular}
\caption{Average time per frame (in seconds) for each of the 3D examples shown in the paper.  Examples were run on workstations with 16-core CPUs running at 2.20 GHz, except for the smoke spheres example, which was run on a cluster equipped with CPUs running at 3.07 GHz and Nvidia Tesla V100 GPUs which were used for the linear solves.}
\label{tbl:perf}
\end{table}

\begin{figure}[!hb]
  \centering
  \begin{subfigure}{.49\columnwidth}
  \includegraphics[draft=\mydraft,trim={0 0 0 0},clip,width=\columnwidth]{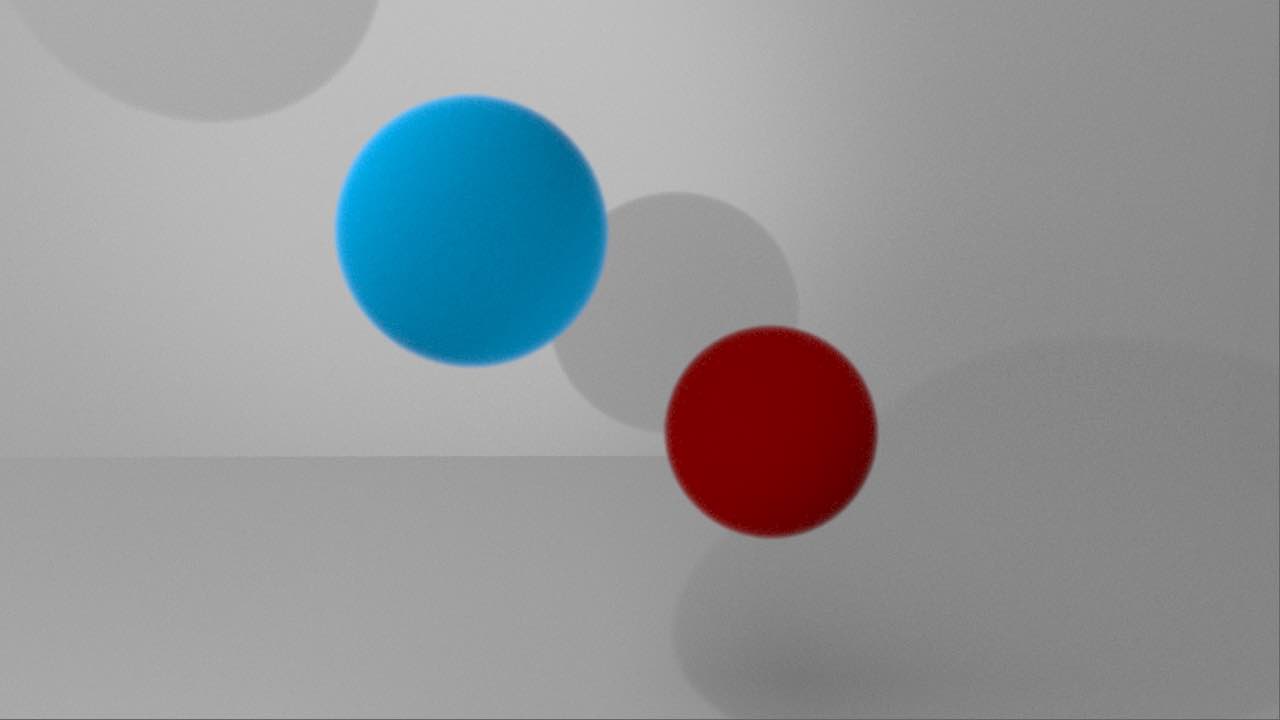}
  \end{subfigure}  %
  \begin{subfigure}{.49\columnwidth}
  \includegraphics[draft=\mydraft,trim={0 0 0 0},clip,width=\columnwidth]{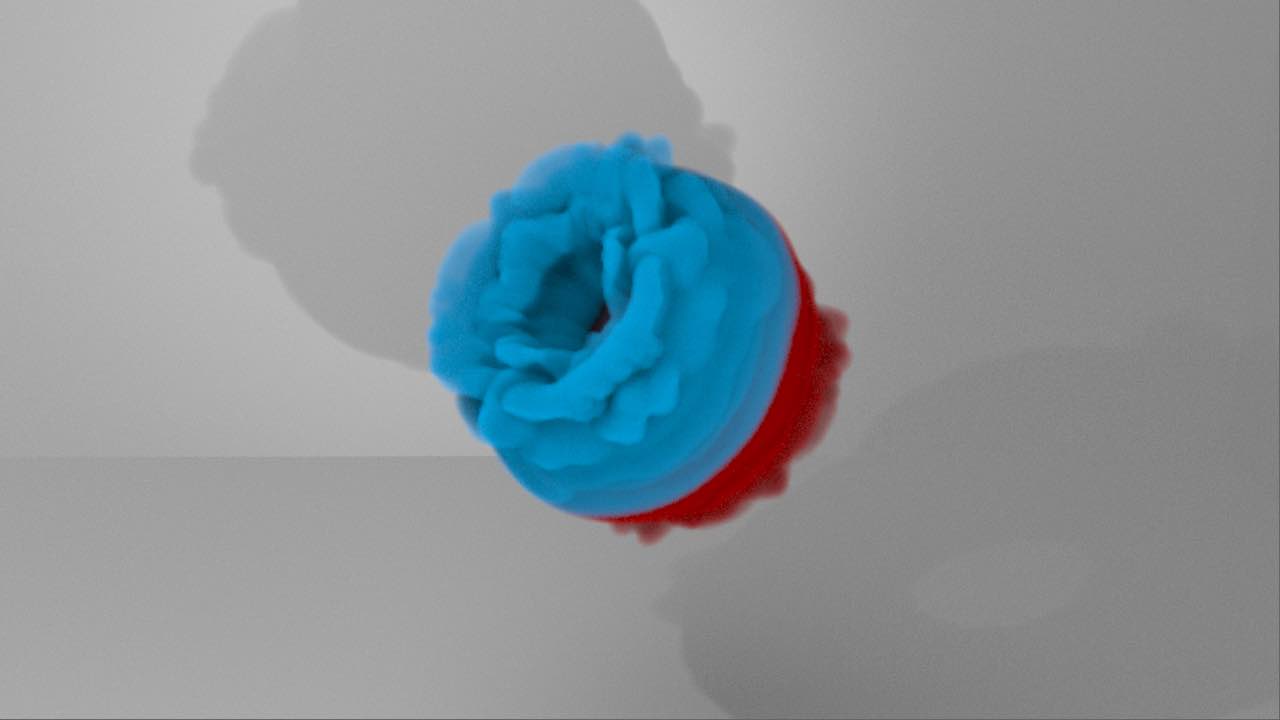}
  \end{subfigure} \\ %
  \begin{subfigure}{.49\columnwidth}
  \includegraphics[draft=\mydraft,trim={0 0 0 0},clip,width=\columnwidth]{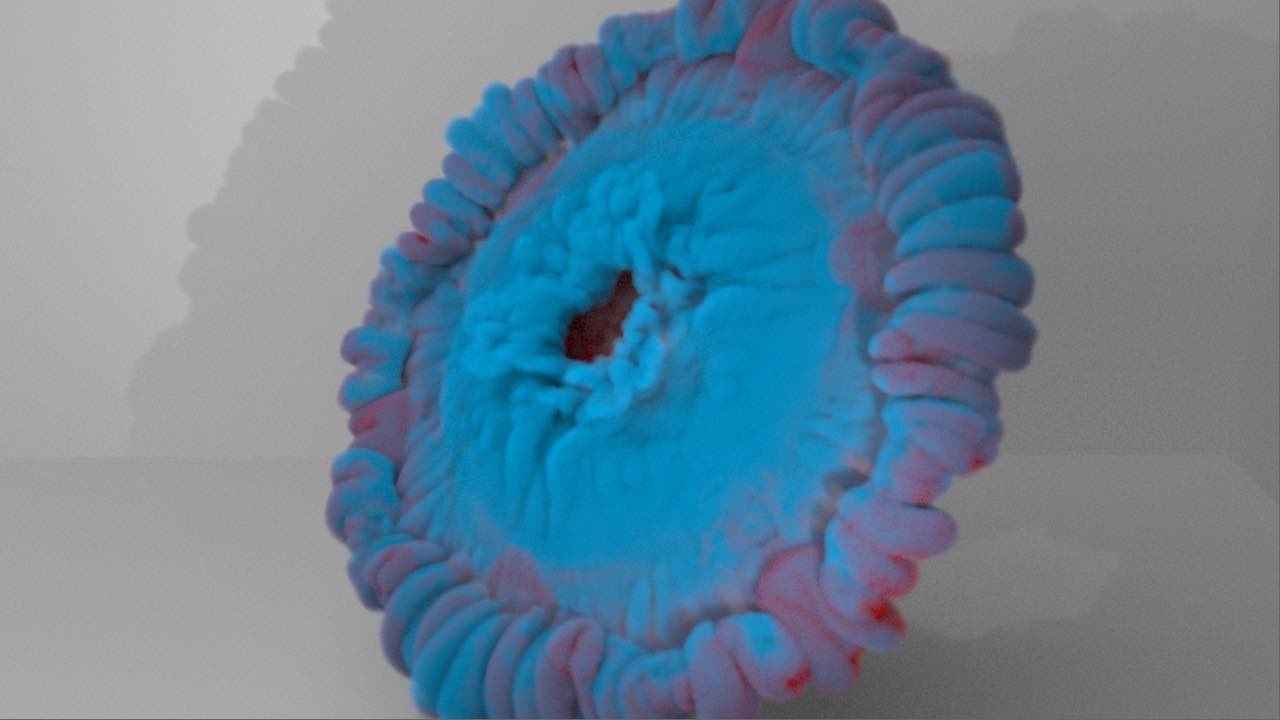}
  \end{subfigure}  %
  \begin{subfigure}{.49\columnwidth}
  \includegraphics[draft=\mydraft,trim={0 0 0 0},clip,width=\columnwidth]{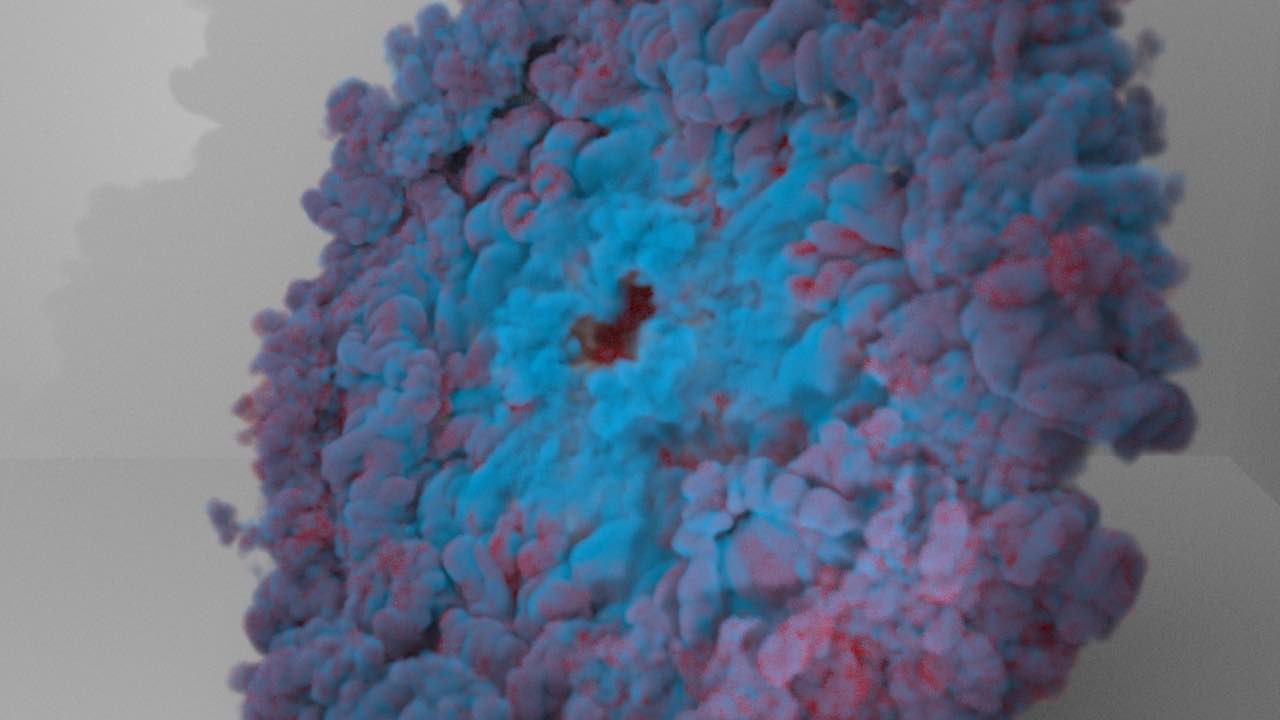}
  \end{subfigure}

\caption{{\textbf{High-resolution smoke}}: Two spheres of smoke collide in a high-resolution 3D simulation ($\Delta x = 1/255$). BSQLB accurately resolves vorticial flow detail.}\label{fig:bigsmoke}
\end{figure}

The implementation of our method takes advantage of hybrid parallelism (MPI, OpenMP, and CUDA/OpenCL) on heterogeneous compute architectures in order to achieve practical runtime performance (see Table~\ref{tbl:perf} for 3D example performance numbers). The spatial domain is uniformly divided into subdomains assigned to distinct MPI ranks, which distributes much of the computational load at the expense of synchronization overhead exchanging ghost information across ranks. On each rank, steps of our time integration loop such as BSLQB advection are multithreaded using OpenMP or CUDA when appropriate. The dominant costs per time step are the solution of the pressure projection system and, in the case of free surface simulation, assembly of the pressure system and its preconditioner. We permute Equation~\eqref{eq:spd_system} so that each rank's degrees of freedom are contiguous in the solution vector then solve the system using AMGCL \cite{demidov:2019:amgcl} using the multi-GPU VexCL backend (or the OpenMP CPU backend on more limited machines). Using a strong algebraic multigrid preconditioner with large-degree Chebyshev smoothing allows our system to be solved to desired tolerance in tens of iterations, even at fine spatial resolution.  An important step in minimizng the cost of system assembly is to scalably parallelize sparse matrix-matrix multiplication, for which we use the algorithm of Saad \shortcite{saad:2003:sparse}.  In the future, we are interested in implementing load balancing strategies such as the simple speculative load balancing approach of \cite{shah:2018:balancing}, particularly for free surface flows. We note that our implementation enables high-resolution simulations such as that in Figure~\ref{fig:bigsmoke} at relatively modest computational cost (see Table~\ref{tbl:perf}).

\section{Discussion and Limitations}
Our approach has several key limitations that could be improved. First, our adoption of collocated multiquadratic velocity and multilinear pressure is a significant departure from most fluid solvers utilized in graphics applications. We note that BSLQB and BSLQB/PolyPIC could be used with a MAC grid; however, each velocity face component would have to be solved for individually. Another drawback for our multiquadratic velocity and multilinear pressure formulation is that it gives rise to a very wide pressure system stencil consisting of 49 non-zero entries per row in 2D and 343 in 3D. Collocated approaches that make use of multilinear velocities and constant pressure give rise to 9 (2D) and 27 (3D) entries per row \cite{zhang:2017:impm}, however they do not allow for $C^1$ continuity and require spurious pressure mode damping. Our wide stencils likely negatively affect the efficacy of preconditioning techniques as well, however we were very pleased with the efficiency of the AMGCL \cite{demidov:2019:amgcl} library. Also, while the use of mass lumping in Equation~\eqref{eq:mass_lump} is necessary to ensure a sparse pressure projection system, Boatella et al. \shortcite{botella:2002:collocation} note that this has been shown to degrade accuracy. In fact, Boatela et al. \shortcite{botella:2002:collocation} introduce a sparse approximate inverse to the full mass matrix to avoid dense systems of equations in the pressure solve without degrading accuracy. Split cubic interpolation, which approximates similar systems with tridiagonal ones could also possibly be used for this \cite{huang:1994:semi}. Adoption of one of these approaches with our formulation would be an interesting area of future work. Also, we note that the more sophisticated transition criteria for narrow banding techniques in Sato et al. \shortcite{sato:2018:nb} could naturally be used with our method.  Finally, we note that the work of Zehnder et al. \shortcite{zehnder:2018:advection,narain:2019:ref} could be easily applied to our technique to further reduce dissipation since it is based on the Chorin \shortcite{chorin:1967:numerical} splitting techniques (Equations~\eqref{eq:split_a}-\eqref{eq:split_div}) that we start from.

\bibliographystyle{ACM-Reference-Format}
\bibliography{paper}

\end{document}